\documentclass[preprint,showpacs,preprintnumbers,amsmath,amssymb]{revtex4}

\usepackage{graphicx}
\usepackage{dcolumn}
\usepackage{bm}

\newcommand{\bov}{\mbox{\boldmath$V$}}
\newcommand{\bop}{\mbox{\boldmath$p$}}
\newcommand{\bbov}{\bar{\mbox{\boldmath$V$}}}
\newcommand{\bbx}{\mbox{\boldmath$x$}}
\newcommand{\boy}{\mbox{\boldmath$y$}}
\newcommand{\boz}{\mbox{\boldmath$z$}}
\newcommand{\boE}{\mbox{\boldmath$E$}}
\newcommand{\boB}{\mbox{\boldmath$B$}}
\newcommand{\bou}{\mbox{\boldmath$u$}}

\begin{document}
\preprint{APS/123-QED}
\title{Structure of a strong supernova shock wave and rapid electron acceleration confined in its transition region}

\author{N. Shimada, M. Hoshino}
\affiliation{Department of Earth and Planetary Science, University of Tokyo, 7-3-1 Hongo, Bunkyo, Tokyo 113-0033, Japan}

\email{nshimada@eps.s.u-tokyo.ac.jp}


\author{T. Amano}
\affiliation{Department of Physics, Nagoya University, Furo-cho, Chikusa, Nagoya, 464-8602, Japan}

\date{\today}


\begin{abstract}

A new rapid energization process within a supernova shock transition region (STR) is reported by utilizing numerical simulation. Although the scale of a STR as a main dissipation region is only several hundreds of thousands km, several interesting structures are found relating to generation of a root of the energetic particles.
The nonlinear evolution of plasma instabilities lead to a dynamical change in the ion phase space distribution which associates with change of the field properties. 
As a result, different types of large-amplitude field structures appear. One is the leading wave packet and another is a series of magnetic solitary humps. Each field structure has a microscopic  scale ($\sim$ the ion inertia length). Through the multiple nonlinear scattering between these large-amplitude field structures, electrons are accelerated directly. Within a STR, quick thermalization realizes energy equipartition between the ion and electron, hot electrons play an important role in keeping these large-amplitude field structures on the ion-acoustic mode. The hot electron shows non-Maxwellian distribution and could be the seed of further non-thermal population.
The "shock system", where fresh incoming and reflected ions are supplied constantly, play an essential role in our result. With a perpendicular shock geometry, the maximum energy of the electron is estimated by equating a width of the STR to a length of the Larmor radius of the energetic electron. Under some realistic condition of $M_A = 170$ and $\omega_{pe}/\Omega_{ce} = 120$, maximum energy is estimated to $\sim$ 10 MeV at an instant only within the STR. 

\end{abstract}

\pacs{Valid PACS appear here}

\maketitle

\section{Introduction}
Recently, the origin of the energetic electron related to astrophysical shocks has been investigated more and more from the point of view of the plasma kinetic processes \cite{SH00, McC01, HS02, Dieck04, Lem04, Dieck08, Mat06, Tr08, AH09}. Within the theory of well-established classical diffusive shock acceleration (DSA) mechanism \cite{Bell87, BE87}, a shock wave tends to be treated as a simple discontinuity. At a shock wave, however, we can find that many kind of plasma kinetics play a key role in the energy dissipation and generation of energetic particles. Correct treatment of the plasma kinetic process around the shock may succeed to settle so-called ``injection problem" to DSA process (e.g. Ref.\cite{Lev96, Ba03, AH07}) as well as to capture a whole stage of DSA process through nonlinear interaction between the shock structure and energetic particles \cite{Bere99, Bell04, ZP08, Cap09}. Furthermore, not only as an assistant role to DSA mechanism, investigation of the energy release mechanism in terms of the plasma kinetics can be a powerful tool to search significant electron energization process in much smaller spatial/time scales than that DSA mechanism may need.

In this paper, we focus on a nonrelativistic ion-electron supernova shock with the perpendicular geometry where the shock normal is perpendicular to the direction of the background magnetic field (e.g. Ref.\cite{Tr08p}). Particle-in-cell (PIC) simulation has been carried out to investigate coupling dynamics among the ion, electron, and field in the shock transition region. The shock transition region (STR) is the front side of the shock wave corresponding to the region flow speed decreases gradually from the upstream toward the shocked downstream region. 
At the high Mach number shocks, part of the incoming ions are always reflected at the shock front and move into the upstream region. In detail, the STR means here the region from the leading edge of the reflected ion to around the end point of the first macroscopic ion gyration in a shocked region. 
The upstream electrons are immediately decelerated just after they meet with the reflected ion. The velocity difference between the decelerated electron and the incoming ion is large enough to excite a series of micro-scale potentials through a strong two-stream instability which grows into a packet of large-amplitude spatial-oscillating magnetic field at the leading edge of the reflected ion (hereafter we call this ``leading packet"). 
When the incoming ion begins to intermingle with the reflected ion, two ion components together make complex vortices in the velocity phase space, the electric and magnetic field structures change immediately. As a result, just after this dynamical change, some components of the leading packet are converted into magnetic solitary humps almost standing in the electron bulk flow frame. 
The electrons bouncing around between the leading packet and the standing magnetic humps get energy rapidly by the 1st-order Fermi type process. 
We define here, the 1st-order Fermi type acceleration as the acceleration of particles by a shock with an extended transition layer, while the original 1st-order Fermi mechanism (DSA) assumes an infinitesimally thin transition layer.
Among magnetic humps, the electrons are also bouncing around and sometimes getting energy in a stochastic manner which depends on mainly local motional electric field caused by a combination of magnetic field of the humps and local bulk flow variation due to the macroscopic ion gyro-motion.
Within the STR, strong thermalization occurs to attain energy equipartition between the electron and ion. The hot electrons contribute to keep above large-amplitude structures on the ion-acoustic mode.
The electron energization mechanisms stated above are quite dynamical process characteristic of a "shock system". 

In the next section we introduce the simulation setup. Section III. presents simulation results: A. macroscopic overview of shock wave properties, B. the energy re-distribution process in the STR, and C. the structure of the STR and characteristic field generation. In subsection III.D we show electron energy spectra and discuss some examples of electron trajectories in the nonlinear field evolution. Section IV. summarizes and discusses our results.

\section{Numerical Simulation Setup}
One-dimensional (1D) electromagnetic relativistic particle-in-cell (PIC) simulation is adopted in the present paper. The shock wave is collisionless and formed by so-called piston method. A high-speed plasma (velocity $u_0$) consisting of the electron and ion (its absolute charge value is equal to the electron) is injected from the left boundary ($x = 0$) and travels toward $x > 0$ region. Initially, a uniform magnetic field $B_z$ (strength $B_0$) perpendicular to the flow direction ($\equiv$ z) is carried by the plasma. The y direction is defined by the relation of each unit vector, $\bbx \times \boy = \boz$. At the right boundary, the plasma flow and all waves are reflected and accumulated to form a shock wave propagating toward the left. The right boundary of the simulation box is taken with enough distance from the left boundary to allow free shock propagation without any artificial downstream disturbances. The initial plasma parameters are as follows: $u_0 / c$ = 0.096, electron and ion beta $\beta_{e} = \beta_{i} = 0.1$ ($\beta_{j}$ = $8\pi T_{j}/B_{0}^{2}$, where $T_j$ is the temperature for $j$ species, $j$ = $e$ for the electron and $j$ = $i$ for the ion), Alfv{\'e}n Mach number ($M_A$) = 174, magnetosonic Mach number ($M_s$) = 159, frequency ratio of the electron plasma oscillation to the electron cyclotron oscillation ($\omega_{pe}/\Omega_{ce}$) is 120, and mass ratio $M/m = 100$. Where $\Omega_{ce}$ = $eB_{0}/mc$, $\omega_{pe}=\sqrt{4\pi n_{0} e^{2}/m}$
 so that $\omega_{pe}/\Omega_{ce} \propto \sqrt{n_0}/B_0$.
The quantities $n_0$, $e$, $M$, $m$, and $c$ are respectively, the upstream density, the electric charge, the ion and electron mass, and the speed of light. 
The electron plasma oscillation time ($2\pi \omega_{pe}^{-1} \equiv T_{pe}$) is divided into 115 numerical time steps. Each simulation cell has 150 particles for each species of the electron and the ion.

We focus here to follow overall dynamics of shock propagation as well as accurate electron dynamics under some realistic plasma parameters, such as high Mach number and the high frequency ratio of $\omega_{pe}/\Omega_{ce}$, with statistically enough particles. Under realistic mass ratio, we require upstream incoming plasma flow of $\sim 6000$ km/s for above setup condition of $\omega_{pe}/\Omega_{ce} = 120$ and $M_A$ = 174 which is a rather strong shock wave but not unrealistic (for example, at RCW 86 shock speed is estimated to 6000 $\pm$ 2800 km/s in Ref.\cite{Hel09}). Although, our simulation has been carried out under 1D and reduced mass ratio condition, results will be one of the useful preparation steps toward complete elucidation of the real supernova shock dynamics. About some two-dimensional effect for the electron dynamics, although in different parameter ranges, there are, for example, \citet{AH09} for non-relativistic case and \citet{Kato07, Spi08} for relativistic case. 

\section{Simulation Results}
\subsection{Overview of shock wave properties}
Figure 1 shows a snapshot of a part of the simulation box in which a shock wave is propagating toward the left. From top, the electron momentum phase space density of x and y direction ($P_{ex}$ and $P_{ey}$), ion momentum phase space density of x and y direction ($P_{ix}$ and $P_{iy}$), electric field $E_x$, and magnetic field $B_z$. These quantities are normalized, respectively, by $m u_0$ for the electron, $M u_0$ for the ion, by the upstream motional electric field $E_0 \equiv u_0 B_0/c$, and by the upstream magnetic field $B_0$. The spatial unit for the horizontal axis is $u_0 / \Omega_{ci}$ ($\equiv L_{STR}$) which is a characteristic scale of the STR, where $\Omega_{ci} = eB_0 / Mc$ is the ion cyclotron frequency. Some arrows labeled (a) $\sim$ (e) will be referred to discuss the electron energy spectra in subsection III.D and Figure 8.
The $E_x$ profile shows large and intense fluctuation generated by microscopic plasma instabilities especially around $x = 0.8 \sim 1.2$ (see subsection III.C in detail). 

Since magnetic field profile looks unfamiliar compared with a classical shock profile, in Figure 2, smoothing profiles (use right tics) are drawn over raw-data profiles (use left tics) with time interval $\sim$ 0.1$\Omega_{ci}^{-1}$. The middle panel is a snapshot at the same time of Figure 1. The time duration from bottom to a top panel corresponds to about 1.4 $T_{STR}$ ($\equiv L_{STR} / u_s$, characteristic STR passage time), where $u_s=1.5 u_0$ is an average incoming flow speed in the shock frame, i.e., shock speed. 
The normalization scales $L_{STR}$ and $T_{STR}$ correspond to 115.2$c/\omega_{pi}$ and 800$\omega_{pi}^{-1}$, respectively, where $\omega_{pi}=\sqrt{4\pi n_{0} e^{2}/M}$ is the ion plasma frequency.
Smoothed profiles with low-resolution are obtained by spatial average over 5 $c/\omega_{pi}$ with each drawing point shifting at an interval of $c/\omega_{pi}$.  Smoothed profiles help to identify familiar shock structure with amplified downstream field about three times as large as upstream magnetic field as expected from Rankin-Hugoniot (R-H) relation (in our 1D case, compression ratio is three). 
As a result of the strong dissipation (thermalization) process as shown in the next subsection, in a low-resolution view, the field profiles and particle distributions (not shown) show little qualitative variation during their propagation \cite{SH05, SM04}. We can discuss properties appear in Figure 1 (normalized simulation time is 7.1) as a typical state of the shock wave. Figures 3-6, 8, and 9 are the same time with Figure 1 and Figures 7, 10, and 11 include data from t = 7.1 $\sim$ 7.7.

\subsection{Rapid energy re-distribution}
Rapid and large velocity spread of the electron is seen at just entrance of the STR ($x = 0.8 \sim 1.0$) in Figure 1. In Figure 3 the quantity of the electron thermalization is shown as the effective temperature, $T_e$ with solid lines compared to the shock flow energy $M u_{s}^2$ (top: linear scale) and to the initial temperature $T_{e0}=T_{i0} \equiv T_0$ (bottom: log scale). The effective temperature is obtained by calculating $\int f(\bop) \gamma m (\bov-\bbov)^2 d{\bop}/\int f(\bop) d{\bop}$, where $f$ is a distribution function with $\bop=\gamma m \bov$, $\gamma$ is the Lorentz factor, $\bov$ is the particle velocity, and $\bbov$ is the averaged particle velocity. The ion temperature ($T_i$) calculated in the same way is also plotted with dotted lines in Figure 3. The local enhancement of the $T_i$ at $x \sim 0.85$ comes from non-Maxwellian behavior with superposition of the incoming and the reflected ion components. When $T_i$ enhancement appears due to the ion reflection, $T_e$ raise up above $T_i$ ($ \ge 10^4 T_{e0}$) within a scale of 0.1$L_{STR}$ (leading edge of a classical foot region). 
In the past, a simulation of mildly relativistic low Mach number shocks ($M_A \sim 2$) under low frequency ratio condition ($\omega_{pe} / \Omega_{ci}<1$) shows quick electron energization, which may realize energy equipartition between ions and electrons (e.g. Ref.\cite{Bessho99}).
To our knowledge, however, such strong electron thermalization observed in the present simulation ($T_e \ge T_i$) has not attained in any other nonrelativistic, lower Mach number shock wave simulations. The remarkable abrupt increase of $T_e$ indicated by arrow at $x\sim 0.92$ corresponds to the position of an onset of the dynamical change caused by the nonlinear saturation of the plasma instability (see next subsection). This means that the nonlinear saturation process plays an important role in the rapid energy equipartition between ions and electrons.
The local hump of $T_e > T_i$ around $x = 1.5 \sim 1.9$ is produced intermittently by a temporary variation of the macroscopic cross-shock potential structure. 
In Figure 4, profiles of $E_x$ (middle, normalized by $E_0$) and the cross-shock potential (bottom, normalized by $M {u_s}^2/2e$) are shown after smoothed over ion-scale (25$c/\omega_{pi}$ for $E_x$, 5$c/\omega_{pi}$ for the potential). Overall variation of $E_x$ is brought by the polarization between the bulk electron and bulk gyrating ion. The bulk electron is overshooting the bulk ion motion. In the top panel of Figure 4, bulk flow variation of the ion (solid), the electron (dashed) are shown with the electron excess amplified 10 times (dotted). Several arrows indicate directions overshooting electrons converge to or diverge to, which causes $E_x$ variation and resultant cross-shock potential. 
The cross-shock potential is one of the fundamental shock structure \cite{Leroy1983} and also a critical quantity in terms of the electron injection efficiency in DSA process \cite{AH07, Baring07}.
Time-averaged cross-shock potential is about $0.3 \sim 0.4 M {u_s}^2/2e$ in our case and consistent to the value used in \citet{AH07}.

Figure 5 illustrates spatial variation of the energy density distribution (\% in a logscale) among the (a) particles, (b) magnetic field $B_z$, and (c) electric field $E_x$. In the panel (a), the ion energy is drawn by dotted line and the electron energy is by solid line. The electron energy is calculated by $(\gamma-1)mc^{2}$ (for ion energy, $m$ is replaced by $M$). Each quantity is normalized by the total energy of all components. Initially, the energy density ratio is about 99\% for the ion and about 1.0\% for the electron. Almost half of the energy initially carried by the incoming ion is converted quit rapidly into the electrons within a scale $< 0.1 L_{STR}$. The maximum energies of $B_z$ and $E_x$ are only of the order $\sim$1\% and $\sim$0.1\%, respectively. Although, the energy carried by the fields is much less than the particle energy, the fields are indispensable especially to the electron energization process as shown below.  

\subsection{Structure of the shock transition region (STR)}
Figure 6 shows an enlargement picture of the entrance of the STR ($x = 0.8 \sim 1.15$) with $P_{ix}$, $E_x$, the electric current of y-component $J_y$ (in a online color version, the total current is black line, the electron current is red line), and $B_z$, from top to bottom. This region can be, roughly speaking, divided into two regions by the vertical line at $x=$0.92. Following the change of the ion phase space distribution, other physical quantities also show clear qualitative change around this line. The left region (we call here ``region I") is where we can easily distinguish the incoming ion (positive velocity component) from the reflected ion (negative velocity component) in the $P_{ix}$ panel. In the region behind the region I (we call here ``region II"), the mixing between the incoming and the reflected ion population progresses rapidly in the phase space. Let us point out some conspicuous structures seen in these regions: the leading packet of the magnetic field (region I) and following magnetic solitary humps with large amplitude electric field oscillation (region II). In the region I, the amplitude of the magnetic field oscillation reaches 10 times as large as the averaged value obtained from R-H relation and in the region II, the amplitude of the electric field oscillation has over 100 times as large as upstream motional electric field ($E_0$). 
In the region I, the nonlinear evolution of the two-stream instability between the decelerated hot electron and the incoming ion \cite{SH04} causes some fluctuation on the ion-acoustic mode. 
As a result, a series of electrostatic potentials are generated which is seen as modulation of the incoming ion as well as the electric field oscillation. Some electrons are trapped around separatrix of these potentials and, consequently, dragged toward $+y$ direction due to motional electric field. Untrapped electrons in non-separatrix region tend to drift toward $-y$ direction under the influence of local $\boE \times \boB$ drift. As a result, a series of clear electric current filaments are formed by the trapped and untrapped electrons (spiky variation from negative to positive in the $J_y$ panel). 
Since the partial derivatives along y and z vanish in our 1D geometry and, in the Ampere's circuital law, the $J_y$ term is more dominant here than the displacement current term $\partial E_y / \partial t$, the relation reduces to $ - |\partial B_z / \partial x | \propto J_y$, which facilitates the spatial modulation of $B_z$ in the leading packet.
A series of resultant negative-positive structures of the magnetic field indicates the possibility of the magnetic reconnection when we carry out 2D or 3D simulations.
The leading packet propagates toward upstream with a velocity nearly equal to the shock speed, while each wavy component of a large-amplitude magnetic field  drifts toward the downstream in the shock frame.

After the electron is heated rapidly and the ion components begin to be diffusive in the region I, another instability between the incoming and reflected ions (ion-ion instability hereafter) evolves. In the region II, mixing between the incoming and the reflected ion component becomes almost complete. The trapped electrons in the region I are detrapped here and clear $J_y$ filaments and negative $B_z$ are disappear in the region II. During this process, some components in the leading packet are modulated into the magnetic humps. 
A possible cause of the magnetic hump formation is evolution and saturation of the ion-ion instability, not general ion-ion ``stream" instability (e.g. \citet{Wu84} in 1D, \citet{OT08} in 2D).
This ion-ion instability looks nonlinear direct coupling between the incoming and reflected ion modulation which is supported by the hot electron on the ion-acoustic mode. The linear analysis shows two-way unstable mode (corresponding to the incoming and reflected ion) with the wavelength $\sim 2 c/ \omega_{pi}$ which is a typical scale of the magnetic humps with rather large growth rate of $20 \sim 30 \Omega_{ci}$ (see Appendix).
Once the magnetic hump structures are formed, they survive over a scale of $\sim L_{STR}$ in the downstream region. Some analytical study about the magnetic humps (magnetosonic solitons), although under much smaller amplitude condition, is reported by \citet{Po07}.

The $P_{ix}$ panel in Figure 6 shows a series of ion phase holes are generated by the ion-ion instability in the region II. In Figure 7, trajectories of the ion phase holes are shown as the propagation of accompanying $E_x$ in a x-time phase space. The two-way complex propagation of the ion phase holes affects electron energization process in a stochastic way. 
These two-way streaming of the ion phase holes can survive Landau damping because the phase speed of the ion phase holes is much smaller than the thermal speed of the hot electron.

In spite of such a high Mach number shock, Buneman instability between the incoming electron and the reflected ion previously reported (e.g. \cite{SH00}) dose not have strong effect on the electron energization here in comparison with other instabilities. The Buneman instability, however, plays an important role to decelerate electrons and make velocity difference from the incoming ion population at the leading edge of the STR. 

When plasma $\beta$ is changed to a larger value, for example, $\sim$ 1.0, we confirm that a structure of the STR is almost identical to $\beta$ $\sim$ 0.1 case here (not shown). The STR structure depends strongly on the separation in the velocity phase space between the incoming and the reflected ion components and on their thermal width, because that affects the way of free energy release via evolution of the plasma kinetic instabilities. At a high-Mach number regime treated here, the velocity separation is large enough to trigger strong nonlinear evolution of the two-stream instability, whichever $\beta$ may be 0.1 $\sim$ 1.

\subsection{Electron energization process}
As seen in Figure 5(a), the electron energy density becomes comparable to the ion energy density instantaneously. How much energy is attained by the most energetic electron at the STR? 
Figure 8 shows spatial evolution of the electron energy spectra around the STR: at just entrance of the STR (a. $x = 0.85 \sim 0.95$), around the ramp region (b. $x = 0.95 \sim 1.1$), around the magnetic overshoot region (c. $x = 1.1 \sim 1.3$), around the magnetic undershoot region (d. $x = 1.3 \sim 1.6$), and its downstream region (e. $x = 1.6 \sim 1.9$). These regions corresponds to the arrows from (a) to (e) in the top of Figure 1. In Figure 8(A) \& (C) the energy is normalized by the injection ion flow energy in the downstream frame ($Mu_0 ^2/2$). The panel (B) shows spectra in terms of $\gamma$. The spectrum becomes harder as goes deeper in the STR. It is surprising that some electrons gain energy more than several times of the incoming ion energy just within the entrance of the STR. 
The most energetic electrons gain energy about 1.5 times larger than the kinetic energy gain due to the ion specular reflection at the shock front. 
Each spectrum shows strong thermalization and a little enhancement of energetic electron tails. In order to compare thermal population to the Maxwellian, we show the energy spectrum at d. with a corresponding Maxwellian of the same effective temperature of the region $x = 1.3 \sim 1.6$ in the panel (C).  
Although we do not find a clear non-thermal tail, some excess is confirmed in higher energy region compared to the Maxwellian. 

In Figure 9, the electron number distribution of different energy ranges are shown from $x = 0.7$ to $x = 1.9$. The top panel shows number of electrons with energy range of $1.5 < \gamma < 1.7$ (solid) and of $\gamma >$ 2.2 (dashed). The middle panel shows population of $\gamma < 1.1$. For reference, the smoothed density profile (normalized by the upstream value) is drawn in the bottom panel. Note that y-axis of the top panel is a log-scale. 
The bulk electron distribution of $\gamma<$1.1 dose not entirely follow the shock-compressed density profile. To keep the shock-compressed density, more energetic electrons are needed to appear behind $x \sim$ 0.9.
The lowest energy electrons ($\gamma <1.1$) shows a little but clear increase with a peak at the leading packet. More energetic electrons in the top panel show exponential rise-up and keep on their flux in the downstream side. The width of the rise-up scale is comparable to the Larmor radius ($r_L$) of the energetic electrons ($r_L = 0.2$ for $\gamma$ = 2.2 electrons, $r_L = 0.11 \sim 0.14$ for $\gamma$ = 1.5 $\sim$ 1.7 electrons). The energetic particles are confined tightly in the STR because the magnetic field has the perpendicular geometry to the shock normal direction. In our simulation, contrary to expectation from DSA mechanism, the energetic electrons are not hovering around over the upstream and downstream regions. 

Next, electron trajectories are investigated. Almost all particles experience both of acceleration and deceleration in a long time duration. We focus here acceleration pattern and pick up two trajectories typical in the leading edge of the STR (Figure 10) and its downstream region (Figure 11). 
In the right panel of Figure 10, the electron trajectory is drawn over time-stacked magnetic field ($B_z$) profile. The strength of $B_z$ is showed by contour bar. A background thermal electron injected into the front of the STR gains energy due to the trapping by the leading packet ($t = 0.0 \sim$0.066) as well as due to bouncing motion between the leading packet and magnetic hump ($t \ge$ 0.066) up to $\gamma \sim$ 3 (about 200 times as large as the initial energy). The kinetic energy gain is brought non-adiabatically. The leading packet is propagating toward upstream as a whole structure (an array of stripes from the lower-right to the upper-left in Figure 10), but each large-amplitude wavy structure is propagating toward downstream (each stripe is toward the upper-right). On the other hand, a series of magnetic humps are almost standing in the electron bulk flow. As a result, the large-amplitude wavy structure is propagating toward the humps. A electron is confined and gains energy through multiple nonlinear interaction with these converging large amplitude $B_z$ structure similar to the 1st-order Fermi type acceleration mechanism.

Figure 11 shows another electron orbit among a series of the magnetic humps in the same format with Figure 10. 
The energization process here is rather mild and stochastic. We confirm that the kinetic energy gain is also brought in a non-adiabatic manner. A rather energetic electron has a large Larmor radius, the effect of the fields can not be symmetric for a round-trip of the electron gyro-motion so that stochastic energy change is brought about.
In Figure 11, the electron shows unmagnetized behavior around time t = 0.05$\sim$0.06 and t = 0.1$\sim$0.15 (indicated by arrows in the right panel) corresponding to the turning point of the gyro-motion. 
The gyro-orbit in a $B_z \sim 0$ field has a quite small curvature so that the electron is going along y-direction for a while. The magnetic mirror force of the humps also keeps the electron in the $B_z \sim 0$ region in this case.
After detrapped from $B_z \sim 0$ region and magnetized, the electron gains some energy due to the motional electric field like so-called ``pick-up ion" process.
After that, the electron is bouncing between humps and gain energy gradually up to $\gamma \sim$4. The nonlinear interaction with these standing magnetic humps do not, however, always bring energy gain. 
It depends not only on the local variation of the electric fields $E_x$ (see the middle panel of Figure 4) but also dominantly on the accumulation of the local $E_y$ variation caused by the bulk flow and $B_z$ variation. 
Around the region shown in Figure 11, there is still positive and negative bulk flow ($\equiv \bou$) variation due to the macroscopic ion gyro-motion as shown in the top of Figure 4. 
The variation of $B_z$ (seen as amplitude variation of the magnetic humps) is connected with the bulk flow variation.  
The gyro-motion of the energetic electrons can be synchronous with the motional electric field $E_y = - e (\pm u) B_z$ along y-direction. As a result, there is sometimes net energy gain through a round-trip of the gyration.

\section{Summary \& Discussion}
The numerical simulation of a collisionless shock wave is reported under high Mach number and high frequency ratio condition ($M_A = 174$ and $\omega_{pe}/\Omega_{ce} = 120$) similar to shock waves generated by the strong blast wave of supernovae. 
We focus on the electron dynamics through the nonlinear particle-field coupling process.
We found that the first rapid electron acceleration occurs in the thin shock front region. Nonlinear evolution of the plasma instabilities between the ions and electrons and following the ion-ion instability causes strong particle energization.
At the moment of the nonlinear saturation of the instability, strong mixing occurs between the incoming and reflected ion component in the velocity phase space as well as some considerable change in the field structures, for example, from the magnetic leading packet to the magnetic solitary hump. 
The leading packet consists of a series of large-amplitude negative-positive magnetic fields. (When a 2D or 3D simulation is carried out, we can expect occurrence of the magnetic reconnection there. It may bring about more interesting electron-field dynamics.)
Since the each component of the leading packet and the magnetic hump are converging, some electrons scattered and bouncing between these two structures gain energy rapidly by the 1st-order Fermi type process.
Some electrons around magnetic humps also gain energy in a rather stochastic manner.
In the magnetic hump region, there is still non-zero bulk flow variation due to the ion gyro-motion.
The electrons gain energy not only due to the local $E_x$ resulted from nonlinear instability evolution, but also due to the motional electric field $E_y$ accompanied by the magnetic hump $B_z$ and bulk flow.
Depending on the gyro-phase, an energetic electron sometimes gains energy when its gyro-motion couples with the $E_y$ variation of the ion-scale.
 
It is surprising that the electron energy density arises rapidly at just an entrance of the STR and becomes comparable to the ion energy density. This energy equipartition is pushed by the nonlinear saturation process of the plasma instability. So-called ``injection problem" in DSA mechanism may be settled if our result remain still true under the realistic dimension and mass ratio condition. We showed that in our simulation, the energetic electrons are confined within a STR while they are interacting with the fields and gaining energy. So that we can estimate the maximum energy $\gamma_{max}$ by equating the electron Larmor radius and characteristic scale of the STR,
\begin{eqnarray}
\displaystyle \frac{mc^2}{eB} \sqrt{\gamma_{max}^2 -1} \sim \displaystyle \frac{1}{2} \frac{\Gamma_0 u_0}{\Omega_{ci}}
\end{eqnarray}
where, $\Gamma_0$ is incoming flow Lorentz factor. Using $u_0 = 2/3 V_A M_A$, above equation results, 
\begin{eqnarray}
\gamma_{max} \sim \displaystyle \sqrt{1+\frac{1}{9} \frac{M}{m}\left( \frac{\omega_{pe}}{\Omega_{ce}} \right) ^{-2} M_A ^{2} \Gamma_0 ^2} 
\end{eqnarray}
When we adopt above equation to the shock wave case in this paper, we have $\gamma_{max} \sim $5 which is consistent to our result of $\gamma_{max} \sim$4. 
With realistic mass ratio $M/m=1836$, $\gamma_{max}$ can reach up to 21 (corresponding to $\sim$ 10 MeV). 
Under a larger mass ratio condition, the STR becomes wider in terms of the scale of electron dynamics and the ion inertia becomes relatively larger. 
Since the ion inertia plays an important role to amplify the fields discussed in the current paper, we can expect more electron energization.
In fact, some periodic simulations in Ref.\cite{SH04} shows that when the mass ratio is larger up to the realistic mass ratio, the electric field supported by the ion inertia becomes more important in the electron energization process.

Recent observation shows shocked temperature $T_e < T_i$ for strong shocks (e.g. Ref.\cite{Rako05, Gh07, He07, Ad08, H09}).
Contrary to the observation, our simulation result shows $T_e \ge T_i$. One reason for that is a scale difference. Our simulation scale is only about several $L_{STR}$ much smaller than observation can resolve, where $L_{STR} \equiv u_0/\Omega_{ci} \sim 4.6\times10^5$ km (for interstellar density of 0.1 cm$^{-3}$), $\sim 1.5\times10^5$ km (for 1.0 cm$^{-3}$) using $\omega_{pe}/\Omega_{ci} = 120$ condition.
The temperature far outside of the STR is, unfortunately, beyond a scope of the paper. Another reason is a shock configuration treated here, namely, the 1D and perpendicular shock system which is a rather strong restriction. For example, 1D simulation tends to overheat electrons compared to the two-dimensional simulations \cite{AH09, Umeda09}.
Although strong thermalization indeed occurs in our simulation, the hot electrons shows non-Maxwellian distribution and non-thermal part grows as leaving off the STR. The hot electrons can be not only a seed of the future non-thermal population but also contribute to further acceleration by keeping the field structures within a scale $\sim L_{STR}$.

Unlike the classical diffusive shock acceleration where the electrons are hovering around shocked and unshocked region due to scattering by large scale MHD waves, the electrons in our simulation experience multiple nonlinear interaction, even within one gyro-motion, with the large-amplitude field structures which generated through the nonlinear evolution of the plasma instabilities. These nonlinear transportations may lead a breakout of the classical diffusion model. 
Recently, without the classical diffusion model, \citet{MD09} also discuss particle acceleration process under multiple nonlinear interaction among a train of magnetic structure ``shocklets" of $c/\omega_{pi}$ scale in the confined region of the STR (shock precursor). As a generator of the structure ``shocklets", nonlinear evolution of the plasma instability is also considered between cosmic ray flow and shock incident flow. Not only for our rather lower energetic regime, also for higher energetic regime, such microscopic coherent structures due to the nonlinear evolution of the plasma instabilities confined in the STR is expected to play a key role in the particle acceleration process.

\acknowledgments
The numerical simulations have been supported by the SX-6 system at ISAS/JAXA, Japan.

\appendix

\section{Linear analysis of the plasma instability}
In order to find linear growth modes of the time scale larger than the electron gyration time, we examine the standard plasma dispersion relation under the magnetized electron and unmagnetized ion condition. As seen in the shock transition region, two ion components of the incoming ions and reflected ions are adopted. We assume that 45\% of the incoming ion is reflected at the shock front (the reflection ratio $\alpha_r$ is 0.45). The relative drift velocity between the two ion components is assumed as 0.5$u_0 (\equiv u_{0d})$. The net electric current is canceled by the total flows of the two ions and electron components. As an initial temperature condition, we take 1$\times 10^3 T_0$ for the incoming ions and 3$\times 10^3 T_0$ for the reflected ions. The magnetic field is assumed to be 3$B_0$. 

When considering waves perpendicular to the magnetic field, the plasma dielectric tensor $\varepsilon(k, \omega)$ is reduced to have the form,
\begin{eqnarray}
\varepsilon_{xx} &=& 1+\sum_n \displaystyle \frac{\omega_{pe}^2}{\omega ^2} \displaystyle \frac{n^2}{\eta} {I_n} \exp (-\eta)\displaystyle \frac{\omega}{n \Omega_{ce}-\omega} 
 + \displaystyle \frac{\omega_{pi0} ^2}{\omega ^2} \zeta_{i0} Z(\zeta_{i0})
 + \displaystyle \frac{\omega_{pir} ^2}{\omega ^2} \zeta_{ir} Z(\zeta_{ir}) \nonumber \\
\varepsilon_{yy} &=& 1+\sum_n \displaystyle \frac{\omega_{pe}^2}{\omega ^2} I_n \exp (-\eta)\displaystyle \frac{\omega}{n \Omega_{ce}-\omega}\nonumber \\
\varepsilon_{xy} &=& i \sum_n \displaystyle \frac{\omega_{pe}^2}{\omega ^2} n({I_n}' - I_n )\exp (-\eta)\displaystyle \frac{\omega}{n \Omega_{ce}-\omega} \nonumber
\end{eqnarray}
where, $k$, $\omega$ are the wave number and wave frequency with the Doppler shift included. The wave frequency in the rest frame ($\omega_r$) is defined by $\omega{_r} = \omega+k u_{0d}(1-2\alpha_r)$. $\omega_{pe}$, $\omega_{pi0}$, $\omega_{pir}$ are the plasma frequencies for the electron, incoming ion, and reflected ion, respectively. $I_n$ and ${I_n}'$ are the modified Bessel function of n-th order with the argument $\eta = v_{e} ^2 k^2/2\Omega_{ce} ^2$ and its derivative. We take the summation for $n$ from $-10$ to 10 here.
$Z$ is the plasma dispersion function with the argument $\zeta_{i0} = (\omega-ku_{0d})/kv_{i0}$ for the incoming ion component and $\zeta_{ir} = (\omega+ku_{0d})/kv_{ir}$ for the reflected ion component. Where $v_{i0}$, $v_{ir}$, and $v_e$ are, respectively, the incoming, reflected, and electron thermal velocity. 
Using above dielectric tensor, the dispersion relation is 
\begin{eqnarray}
\varepsilon_{xx} \left[ \varepsilon_{yy} - \displaystyle \left( \frac{ck}{\omega} \right)^2 \right] + \varepsilon_{xy}^2 = 0
\end{eqnarray}

Figure 12 shows some of the solutions of eq.(A1). Top panels include real part of $\omega$ and bottom panels include imaginary part of $\omega$ ($\equiv \gamma$, growth rate) versus wavelength $\lambda$. The panels (a) and (b) belongs to the incoming ion because their phase velocity is the same direction with the incoming ion. The panels (c) and (d) belongs to the reflected ion with the same reason. Their harmonics in other wave number range is not shown. $\omega$ and $\lambda$ are normalized by the plasma frequency of the upstream ion $\omega_{pi}$ and its inertia $c/\omega_{pi}$, respectively. The electron temperature $T_e$ is set up to 2$\times 10^3 T_0$ (dotted), 4$\times 10^3 T_0$ (long-dotted), 8$\times 10^3 T_0$ (dashed), and 1$\times 10^4 T_0$ (solid).
As the electron temperature rises up, the growth rate increases with its peak moves a little toward smaller $\lambda$. The solution does not have a strong dependence on the ion temperature but smaller ion temperature results larger growth rate. The wavelength $\sim$ 2$c/\omega_{pi}$ (comparable to the normalized width $\Delta x \sim 0.01$) around the peak of the growth rate corresponds to the width of the magnetic hump stated in subsection III.C. Coupling between such linear growth modes in the incoming and reflected ion component can be some cause of the magnetic hump generation. However, since strong nonlinearity is observed in the simulation, especially in region II, we can not separate clearly the linear growth from other nonlinear plasma actions.

\newpage 

\begin{thebibliography}{38}
\expandafter\ifx\csname natexlab\endcsname\relax\def\natexlab#1{#1}\fi
\expandafter\ifx\csname bibnamefont\endcsname\relax
  \def\bibnamefont#1{#1}\fi
\expandafter\ifx\csname bibfnamefont\endcsname\relax
  \def\bibfnamefont#1{#1}\fi
\expandafter\ifx\csname citenamefont\endcsname\relax
  \def\citenamefont#1{#1}\fi
\expandafter\ifx\csname url\endcsname\relax
  \def\url#1{\texttt{#1}}\fi
\expandafter\ifx\csname urlprefix\endcsname\relax\def\urlprefix{URL }\fi
\providecommand{\bibinfo}[2]{#2}
\providecommand{\eprint}[2][]{\url{#2}}

\bibitem[{\citenamefont{Shimada and Hoshino}(2000)}]{SH00}
\bibinfo{author}{\bibfnamefont{N.}~\bibnamefont{Shimada}} \bibnamefont{and}
  \bibinfo{author}{\bibfnamefont{M.}~\bibnamefont{Hoshino}},
  \bibinfo{journal}{Astrophys. J. Lett.} \textbf{\bibinfo{volume}{543}},
  \bibinfo{pages}{L67} (\bibinfo{year}{2000}).

\bibitem[{\citenamefont{McClements et~al.}(2001)\citenamefont{McClements,
  Dieckmann, Ynnerman, Chapman, and Dendy}}]{McC01}
\bibinfo{author}{\bibfnamefont{K.~G.} \bibnamefont{McClements}},
  \bibinfo{author}{\bibfnamefont{M.~E.} \bibnamefont{Dieckmann}},
  \bibinfo{author}{\bibfnamefont{A.}~\bibnamefont{Ynnerman}},
  \bibinfo{author}{\bibfnamefont{S.~C.} \bibnamefont{Chapman}},
  \bibnamefont{and} \bibinfo{author}{\bibfnamefont{R.~O.} \bibnamefont{Dendy}},
  \bibinfo{journal}{Phys. Rev. Lett.} \textbf{\bibinfo{volume}{87}},
  \bibinfo{pages}{255002} (\bibinfo{year}{2001}).

\bibitem[{\citenamefont{Hoshino and Shimada}(2002)}]{HS02}
\bibinfo{author}{\bibfnamefont{M.}~\bibnamefont{Hoshino}} \bibnamefont{and}
  \bibinfo{author}{\bibfnamefont{N.}~\bibnamefont{Shimada}},
  \bibinfo{journal}{Astrophys. J.} \textbf{\bibinfo{volume}{572}},
  \bibinfo{pages}{880} (\bibinfo{year}{2002}).

\bibitem[{\citenamefont{Dieckmann et~al.}(2004)\citenamefont{Dieckmann,
  Eliasson, Stathopoulos, and Ynnerman}}]{Dieck04}
\bibinfo{author}{\bibfnamefont{M.~E.} \bibnamefont{Dieckmann}},
  \bibinfo{author}{\bibfnamefont{B.}~\bibnamefont{Eliasson}},
  \bibinfo{author}{\bibfnamefont{A.}~\bibnamefont{Stathopoulos}},
  \bibnamefont{and} \bibinfo{author}{\bibfnamefont{A.}~\bibnamefont{Ynnerman}},
  \bibinfo{journal}{\prl} \textbf{\bibinfo{volume}{92}},
  \bibinfo{pages}{065006} (\bibinfo{year}{2004}).

\bibitem[{\citenamefont{Lembege et~al.}(2004)\citenamefont{Lembege, Giacalone,
  Scholer, Hada, Hoshino, Krasnoselskikh, Kucharek, Savoini, and
  Terasawa}}]{Lem04}
\bibinfo{author}{\bibfnamefont{B.}~\bibnamefont{Lembege}},
  \bibinfo{author}{\bibfnamefont{J.}~\bibnamefont{Giacalone}},
  \bibinfo{author}{\bibfnamefont{M.}~\bibnamefont{Scholer}},
  \bibinfo{author}{\bibfnamefont{T.}~\bibnamefont{Hada}},
  \bibinfo{author}{\bibfnamefont{M.}~\bibnamefont{Hoshino}},
  \bibinfo{author}{\bibfnamefont{V.}~\bibnamefont{Krasnoselskikh}},
  \bibinfo{author}{\bibfnamefont{H.}~\bibnamefont{Kucharek}},
  \bibinfo{author}{\bibfnamefont{P.}~\bibnamefont{Savoini}}, \bibnamefont{and}
  \bibinfo{author}{\bibfnamefont{T.}~\bibnamefont{Terasawa}},
  \bibinfo{journal}{Space Sci. Rev.} \textbf{\bibinfo{volume}{110}},
  \bibinfo{pages}{161} (\bibinfo{year}{2004}).

\bibitem[{\citenamefont{Dieckmann et~al.}(2008)\citenamefont{Dieckmann, Shukla,
  and Drury}}]{Dieck08}
\bibinfo{author}{\bibfnamefont{M.~E.} \bibnamefont{Dieckmann}},
  \bibinfo{author}{\bibfnamefont{P.~K.} \bibnamefont{Shukla}},
  \bibnamefont{and} \bibinfo{author}{\bibfnamefont{L.~O.~C.}
  \bibnamefont{Drury}}, \bibinfo{journal}{\apj} \textbf{\bibinfo{volume}{675}},
  \bibinfo{pages}{586} (\bibinfo{year}{2008}).

\bibitem[{\citenamefont{Matsukiyo and Scholer}(2006)}]{Mat06}
\bibinfo{author}{\bibfnamefont{S.}~\bibnamefont{Matsukiyo}} \bibnamefont{and}
  \bibinfo{author}{\bibfnamefont{M.}~\bibnamefont{Scholer}},
  \bibinfo{journal}{J. Geophys.} \textbf{\bibinfo{volume}{111}},
  \bibinfo{pages}{CiteID A06104} (\bibinfo{year}{2006}).

\bibitem[{\citenamefont{Treumann and Jaroschek}(2008{\natexlab{a}})}]{Tr08}
\bibinfo{author}{\bibfnamefont{R.~A.} \bibnamefont{Treumann}} \bibnamefont{and}
  \bibinfo{author}{\bibfnamefont{C.~H.} \bibnamefont{Jaroschek}},
  \bibinfo{journal}{arXiv:0806.4046 Fundamentals of Non-relativistic
  Collisionless Shock Physics: V. Acceleration of Charged Particles (Astron.
  and Astrophys. Rev.)}  (\bibinfo{year}{2008}{\natexlab{a}}).

\bibitem[{\citenamefont{Amano and Hoshino}(2009)}]{AH09}
\bibinfo{author}{\bibfnamefont{T.}~\bibnamefont{Amano}} \bibnamefont{and}
  \bibinfo{author}{\bibfnamefont{M.}~\bibnamefont{Hoshino}},
  \bibinfo{journal}{\apj} \textbf{\bibinfo{volume}{690}}, \bibinfo{pages}{244}
  (\bibinfo{year}{2009}).

\bibitem[{\citenamefont{Bell}(1987)}]{Bell87}
\bibinfo{author}{\bibfnamefont{A.~R.} \bibnamefont{Bell}},
  \bibinfo{journal}{Mon. Not. R. Astron. Soc.} \textbf{\bibinfo{volume}{225}},
  \bibinfo{pages}{615} (\bibinfo{year}{1987}).

\bibitem[{\citenamefont{Blandford and Eichler}(1987)}]{BE87}
\bibinfo{author}{\bibfnamefont{R.}~\bibnamefont{Blandford}} \bibnamefont{and}
  \bibinfo{author}{\bibfnamefont{D.}~\bibnamefont{Eichler}},
  \bibinfo{journal}{Phys. Rep.} \textbf{\bibinfo{volume}{154}},
  \bibinfo{pages}{1} (\bibinfo{year}{1987}).

\bibitem[{\citenamefont{Levinson}(1996)}]{Lev96}
\bibinfo{author}{\bibfnamefont{A.}~\bibnamefont{Levinson}},
  \bibinfo{journal}{Mon. Not. R. Astron. Soc.} \textbf{\bibinfo{volume}{278}},
  \bibinfo{pages}{1018} (\bibinfo{year}{1996}).

\bibitem[{\citenamefont{A.~Bamba et~al.}(2003)\citenamefont{A.~Bamba, Yamazaki,
  Ueno, and Koyama}}]{Ba03}
\bibinfo{author}{\bibfnamefont{A.}~\bibnamefont{A.~Bamba}},
  \bibinfo{author}{\bibfnamefont{R.}~\bibnamefont{Yamazaki}},
  \bibinfo{author}{\bibfnamefont{M.}~\bibnamefont{Ueno}}, \bibnamefont{and}
  \bibinfo{author}{\bibfnamefont{K.}~\bibnamefont{Koyama}},
  \bibinfo{journal}{\apj} \textbf{\bibinfo{volume}{589}}, \bibinfo{pages}{827}
  (\bibinfo{year}{2003}).

\bibitem[{\citenamefont{Amano and Hoshino}(2007)}]{AH07}
\bibinfo{author}{\bibfnamefont{T.}~\bibnamefont{Amano}} \bibnamefont{and}
  \bibinfo{author}{\bibfnamefont{M.}~\bibnamefont{Hoshino}},
  \bibinfo{journal}{\apj} \textbf{\bibinfo{volume}{661}}, \bibinfo{pages}{190}
  (\bibinfo{year}{2007}).

\bibitem[{\citenamefont{Berezhko and Ellison}(1999)}]{Bere99}
\bibinfo{author}{\bibfnamefont{E.~G.} \bibnamefont{Berezhko}} \bibnamefont{and}
  \bibinfo{author}{\bibfnamefont{D.~C.} \bibnamefont{Ellison}},
  \bibinfo{journal}{\apj} \textbf{\bibinfo{volume}{526}}, \bibinfo{pages}{385}
  (\bibinfo{year}{1999}).

\bibitem[{\citenamefont{Bell}(2004)}]{Bell04}
\bibinfo{author}{\bibfnamefont{A.~R.} \bibnamefont{Bell}},
  \bibinfo{journal}{Mon. Not. R. Astron. Soc.} \textbf{\bibinfo{volume}{353}},
  \bibinfo{pages}{550} (\bibinfo{year}{2004}).

\bibitem[{\citenamefont{Zirakashvili and Ptuskin}(2008)}]{ZP08}
\bibinfo{author}{\bibfnamefont{V.~N.} \bibnamefont{Zirakashvili}}
  \bibnamefont{and} \bibinfo{author}{\bibfnamefont{V.~S.}
  \bibnamefont{Ptuskin}}, \bibinfo{journal}{\apj}
  \textbf{\bibinfo{volume}{678}}, \bibinfo{pages}{939} (\bibinfo{year}{2008}).

\bibitem[{\citenamefont{Caprioli et~al.}(2009)\citenamefont{Caprioli, Blasi,
  Amato, and Vietri}}]{Cap09}
\bibinfo{author}{\bibfnamefont{D.}~\bibnamefont{Caprioli}},
  \bibinfo{author}{\bibfnamefont{P.}~\bibnamefont{Blasi}},
  \bibinfo{author}{\bibfnamefont{E.}~\bibnamefont{Amato}}, \bibnamefont{and}
  \bibinfo{author}{\bibfnamefont{M.}~\bibnamefont{Vietri}},
  \bibinfo{journal}{Mon. Not. R. Astron. Soc.} \textbf{\bibinfo{volume}{395}},
  \bibinfo{pages}{895} (\bibinfo{year}{2009}).

\bibitem[{\citenamefont{Treumann and Jaroschek}(2008{\natexlab{b}})}]{Tr08p}
\bibinfo{author}{\bibfnamefont{R.~A.} \bibnamefont{Treumann}} \bibnamefont{and}
  \bibinfo{author}{\bibfnamefont{C.~H.} \bibnamefont{Jaroschek}},
  \bibinfo{journal}{arXiv:0805.2181 Fundamentals of Non-relativistic
  Collisionless Shock Physics: III. Quasi-Perpendicular Supercritical Shocks
  (Astron. and Astrophys. Rev.)}  (\bibinfo{year}{2008}{\natexlab{b}}).

\bibitem[{\citenamefont{Helder et~al.}(2009)\citenamefont{Helder, Vink, Bassa,
  Bamba, Bleeker, Funk, Ghavamian, van~der Heyden, Verbunt, and
  Yamazaki}}]{Hel09}
\bibinfo{author}{\bibfnamefont{E.~A.} \bibnamefont{Helder}},
  \bibinfo{author}{\bibfnamefont{J.}~\bibnamefont{Vink}},
  \bibinfo{author}{\bibfnamefont{.~G.} \bibnamefont{Bassa}},
  \bibinfo{author}{\bibfnamefont{A.}~\bibnamefont{Bamba}},
  \bibinfo{author}{\bibfnamefont{J.~A.~M.} \bibnamefont{Bleeker}},
  \bibinfo{author}{\bibfnamefont{S.}~\bibnamefont{Funk}},
  \bibinfo{author}{\bibfnamefont{P.}~\bibnamefont{Ghavamian}},
  \bibinfo{author}{\bibfnamefont{K.~J.} \bibnamefont{van~der Heyden}},
  \bibinfo{author}{\bibfnamefont{F.}~\bibnamefont{Verbunt}}, \bibnamefont{and}
  \bibinfo{author}{\bibfnamefont{R.}~\bibnamefont{Yamazaki}},
  \bibinfo{journal}{Science} \textbf{\bibinfo{volume}{325}},
  \bibinfo{pages}{719} (\bibinfo{year}{2009}).

\bibitem[{\citenamefont{Kato}(2007)}]{Kato07}
\bibinfo{author}{\bibfnamefont{T.~N.} \bibnamefont{Kato}},
  \bibinfo{journal}{\apj} \textbf{\bibinfo{volume}{668}}, \bibinfo{pages}{974}
  (\bibinfo{year}{2007}).

\bibitem[{\citenamefont{Spitkovsky}(2008)}]{Spi08}
\bibinfo{author}{\bibfnamefont{A.}~\bibnamefont{Spitkovsky}},
  \bibinfo{journal}{\apj} \textbf{\bibinfo{volume}{673}}, \bibinfo{pages}{L39}
  (\bibinfo{year}{2008}).

\bibitem[{\citenamefont{Shimada and Hoshino}(2005)}]{SH05}
\bibinfo{author}{\bibfnamefont{N.}~\bibnamefont{Shimada}} \bibnamefont{and}
  \bibinfo{author}{\bibfnamefont{M.}~\bibnamefont{Hoshino}},
  \bibinfo{journal}{J. Geophys.} \textbf{\bibinfo{volume}{110}},
  \bibinfo{pages}{CiteID A02105} (\bibinfo{year}{2005}).

\bibitem[{\citenamefont{Scholer and Matsukiyo}(2004)}]{SM04}
\bibinfo{author}{\bibfnamefont{M.}~\bibnamefont{Scholer}} \bibnamefont{and}
  \bibinfo{author}{\bibfnamefont{S.}~\bibnamefont{Matsukiyo}},
  \bibinfo{journal}{Ann. Geophys.} \textbf{\bibinfo{volume}{22}},
  \bibinfo{pages}{2345} (\bibinfo{year}{2004}).

\bibitem[{\citenamefont{Bessho and Ohsawa}(1999)}]{Bessho99}
\bibinfo{author}{\bibfnamefont{N.}~\bibnamefont{Bessho}} \bibnamefont{and}
  \bibinfo{author}{\bibfnamefont{Y.}~\bibnamefont{Ohsawa}},
  \bibinfo{journal}{Phys. Plasmas} \textbf{\bibinfo{volume}{6}},
  \bibinfo{pages}{3076} (\bibinfo{year}{1999}).

\bibitem[{\citenamefont{Leroy}(1983)}]{Leroy1983}
\bibinfo{author}{\bibfnamefont{M.~M.} \bibnamefont{Leroy}},
  \bibinfo{journal}{Phys. Fluids} \textbf{\bibinfo{volume}{26}},
  \bibinfo{pages}{2742} (\bibinfo{year}{1983}).

\bibitem[{\citenamefont{Baring and Summerlin}(2007)}]{Baring07}
\bibinfo{author}{\bibfnamefont{M.~G.} \bibnamefont{Baring}} \bibnamefont{and}
  \bibinfo{author}{\bibfnamefont{E.~J.} \bibnamefont{Summerlin}},
  \bibinfo{journal}{Astrophys. Space Sci.} \textbf{\bibinfo{volume}{307}},
  \bibinfo{pages}{165} (\bibinfo{year}{2007}).

\bibitem[{\citenamefont{Shimada and Hoshino}(2004)}]{SH04}
\bibinfo{author}{\bibfnamefont{N.}~\bibnamefont{Shimada}} \bibnamefont{and}
  \bibinfo{author}{\bibfnamefont{M.}~\bibnamefont{Hoshino}},
  \bibinfo{journal}{Phys. Plasmas} \textbf{\bibinfo{volume}{11}},
  \bibinfo{pages}{1840} (\bibinfo{year}{2004}).

\bibitem[{\citenamefont{Wu et~al.}(1984)\citenamefont{Wu, Winske, Tanaka,
  Papadopoulos, Akimoto, Goodrich, Zhou, M.Tsai, T., Rodriguez et~al.}}]{Wu84}
\bibinfo{author}{\bibfnamefont{C.~S.} \bibnamefont{Wu}},
  \bibinfo{author}{\bibfnamefont{D.}~\bibnamefont{Winske}},
  \bibinfo{author}{\bibfnamefont{M.}~\bibnamefont{Tanaka}},
  \bibinfo{author}{\bibfnamefont{K.}~\bibnamefont{Papadopoulos}},
  \bibinfo{author}{\bibfnamefont{K.}~\bibnamefont{Akimoto}},
  \bibinfo{author}{\bibfnamefont{C.~C.} \bibnamefont{Goodrich}},
  \bibinfo{author}{\bibfnamefont{Y.~M.} \bibnamefont{Zhou}},
  \bibinfo{author}{\bibfnamefont{Y.}~\bibnamefont{M.Tsai}},
  \bibinfo{author}{\bibfnamefont{S.}~\bibnamefont{T.}},
  \bibinfo{author}{\bibfnamefont{P.}~\bibnamefont{Rodriguez}},
  \bibnamefont{et~al.}, \bibinfo{journal}{Space Sci. Rev.}
  \textbf{\bibinfo{volume}{37}}, \bibinfo{pages}{63} (\bibinfo{year}{1984}).

\bibitem[{\citenamefont{Ohira and Takahara}(2008)}]{OT08}
\bibinfo{author}{\bibfnamefont{Y.}~\bibnamefont{Ohira}} \bibnamefont{and}
  \bibinfo{author}{\bibfnamefont{F.}~\bibnamefont{Takahara}},
  \bibinfo{journal}{\apj} \textbf{\bibinfo{volume}{688}}, \bibinfo{pages}{320}
  (\bibinfo{year}{2008}).

\bibitem[{\citenamefont{Pokhotelov et~al.}(2007)\citenamefont{Pokhotelov,
  Balikhin, Onishchenko, and Walker}}]{Po07}
\bibinfo{author}{\bibfnamefont{O.~A.} \bibnamefont{Pokhotelov}},
  \bibinfo{author}{\bibfnamefont{M.~A.} \bibnamefont{Balikhin}},
  \bibinfo{author}{\bibfnamefont{O.~G.} \bibnamefont{Onishchenko}},
  \bibnamefont{and} \bibinfo{author}{\bibfnamefont{S.~N.}
  \bibnamefont{Walker}}, \bibinfo{journal}{Planetary, Space Sci.}
  \textbf{\bibinfo{volume}{55}}, \bibinfo{pages}{2310} (\bibinfo{year}{2007}).

\bibitem[{\citenamefont{Rakowski}(2005)}]{Rako05}
\bibinfo{author}{\bibfnamefont{C.~E.} \bibnamefont{Rakowski}},
  \bibinfo{journal}{Adv. Space Res} \textbf{\bibinfo{volume}{35}},
  \bibinfo{pages}{1017} (\bibinfo{year}{2005}).

\bibitem[{\citenamefont{Ghavamian et~al.}(2007)\citenamefont{Ghavamian, Laming,
  and Rakowski}}]{Gh07}
\bibinfo{author}{\bibfnamefont{P.}~\bibnamefont{Ghavamian}},
  \bibinfo{author}{\bibfnamefont{J.~M.} \bibnamefont{Laming}},
  \bibnamefont{and} \bibinfo{author}{\bibfnamefont{C.~E.}
  \bibnamefont{Rakowski}}, \bibinfo{journal}{\apj}
  \textbf{\bibinfo{volume}{654}}, \bibinfo{pages}{L69} (\bibinfo{year}{2007}).

\bibitem[{\citenamefont{Heng and McCray}(2007)}]{He07}
\bibinfo{author}{\bibfnamefont{K.}~\bibnamefont{Heng}} \bibnamefont{and}
  \bibinfo{author}{\bibfnamefont{R.}~\bibnamefont{McCray}},
  \bibinfo{journal}{\apj} \textbf{\bibinfo{volume}{654}}, \bibinfo{pages}{923}
  (\bibinfo{year}{2007}).

\bibitem[{\citenamefont{van Adelsberg et~al.}(2008)\citenamefont{van Adelsberg,
  Heng, McCray, and Raymond}}]{Ad08}
\bibinfo{author}{\bibfnamefont{M.}~\bibnamefont{van Adelsberg}},
  \bibinfo{author}{\bibfnamefont{K.}~\bibnamefont{Heng}},
  \bibinfo{author}{\bibfnamefont{R.}~\bibnamefont{McCray}}, \bibnamefont{and}
  \bibinfo{author}{\bibfnamefont{J.~C.} \bibnamefont{Raymond}},
  \bibinfo{journal}{\apj} \textbf{\bibinfo{volume}{689}}, \bibinfo{pages}{1089}
  (\bibinfo{year}{2008}).

\bibitem[{\citenamefont{Heng}(2009)}]{H09}
\bibinfo{author}{\bibfnamefont{K.}~\bibnamefont{Heng}},
  \bibinfo{journal}{arXiv0908.4080H Balmer-Dominated Shocks: A Concise Review
  (Pub. of the Astronomical Soc. of Australia)}  (\bibinfo{year}{2009}).

\bibitem[{\citenamefont{Umeda et~al.}(2009)\citenamefont{Umeda, Yamao, and
  Yamazaki}}]{Umeda09}
\bibinfo{author}{\bibfnamefont{T.}~\bibnamefont{Umeda}},
  \bibinfo{author}{\bibfnamefont{M.}~\bibnamefont{Yamao}}, \bibnamefont{and}
  \bibinfo{author}{\bibfnamefont{R.}~\bibnamefont{Yamazaki}},
  \bibinfo{journal}{\apj} \textbf{\bibinfo{volume}{695}}, \bibinfo{pages}{574}
  (\bibinfo{year}{2009}).

\bibitem[{\citenamefont{Malkov and Diamond}(2009)}]{MD09}
\bibinfo{author}{\bibfnamefont{M.~A.} \bibnamefont{Malkov}} \bibnamefont{and}
  \bibinfo{author}{\bibfnamefont{P.~H.} \bibnamefont{Diamond}},
  \bibinfo{journal}{\apj} \textbf{\bibinfo{volume}{692}}, \bibinfo{pages}{1571}
  (\bibinfo{year}{2009}).

\end{thebibliography}

\newpage

\clearpage

\begin{figure} 
\includegraphics[scale=0.7]{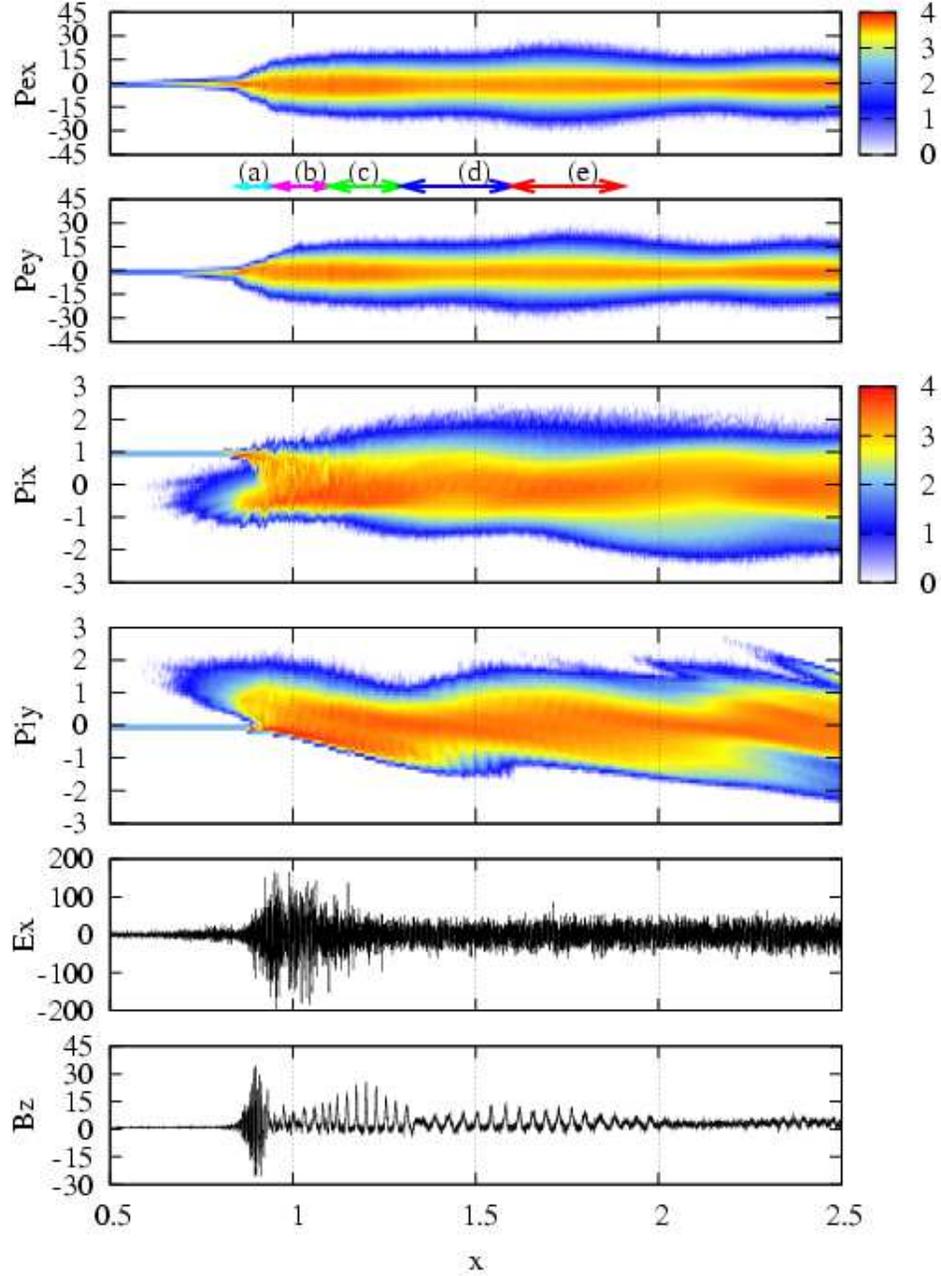}
\caption{(Color online) The structure of a high Mach number shock wave. From top, the electron momentum of $x$ and $y$ components ($P_{ex}$ and $P_{ey}$), the ion momentum of $x$ and $y$ components ($P_{ix}$ and $P_{iy}$), the electric field ($E_x$), and the magnetic field ($B_z$). The contour shows 10 logarithmic distribution of the particle number. The electron and ion momenta are normalized by $m u_0$ and $M u_0$, respectively, the electric and magnetic fields are normalized by the upstream motional electric field $E_0 \equiv u_0 B_0/c$, and upstream magnetic field $B_0$, respectively. The horizontal axis is normalized by $u_0 / \Omega_{ci}$. Several arrows (a)-(e) are referred in Figure 8. The normalized simulation time for the figure is 7.1.}
\end{figure}

\clearpage
\begin{figure} 
\includegraphics{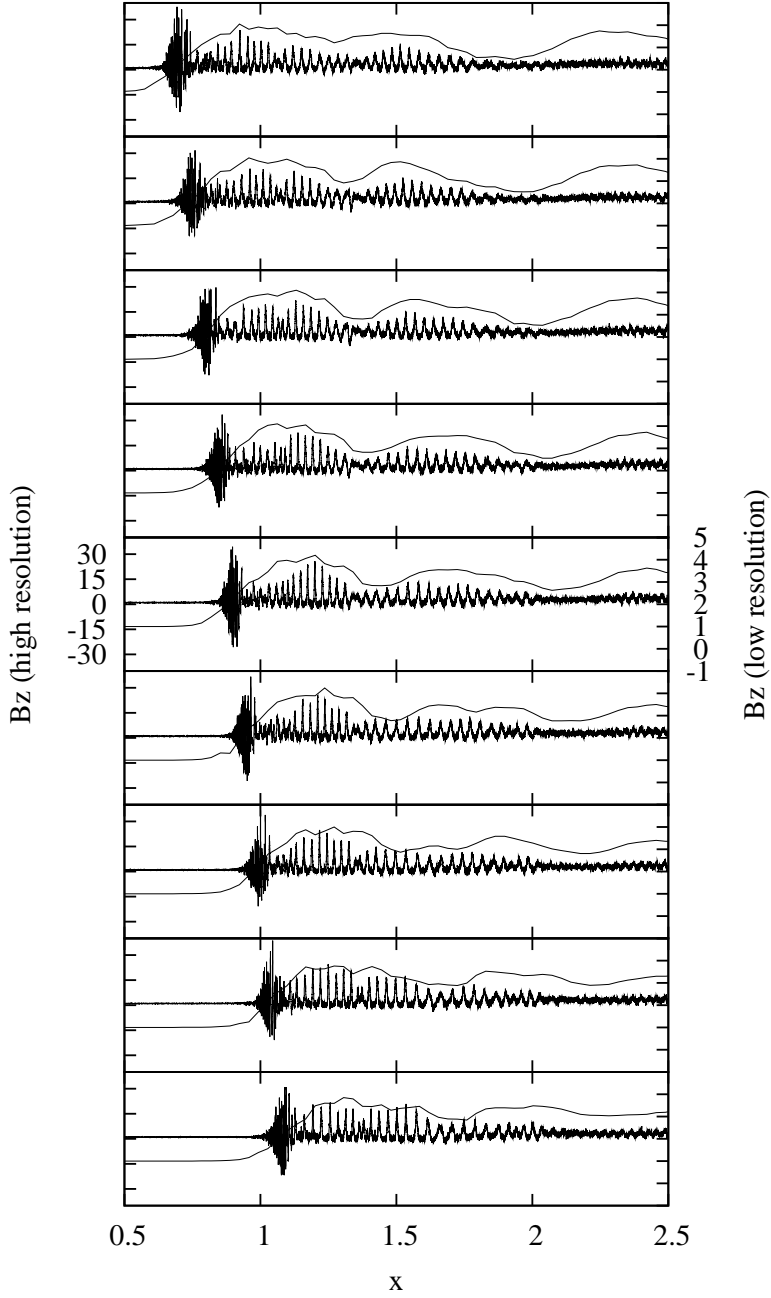}
\caption{$B_z$ profiles with raw value (left tics) and with smoothing value (right tics) are drawn at an interval of $\sim$ 0.1$\Omega_{ci}^{-1}$ from bottom to a top panel. Smoothed profiles are obtained by spatial average over 5$c/\omega_{pi}$ with each drawing point shifted at an interval of $c/\omega_{pi}$. The time of the middle panel corresponds to the time of Figure 1.}
\end{figure}

\clearpage
\begin{figure} 
\includegraphics{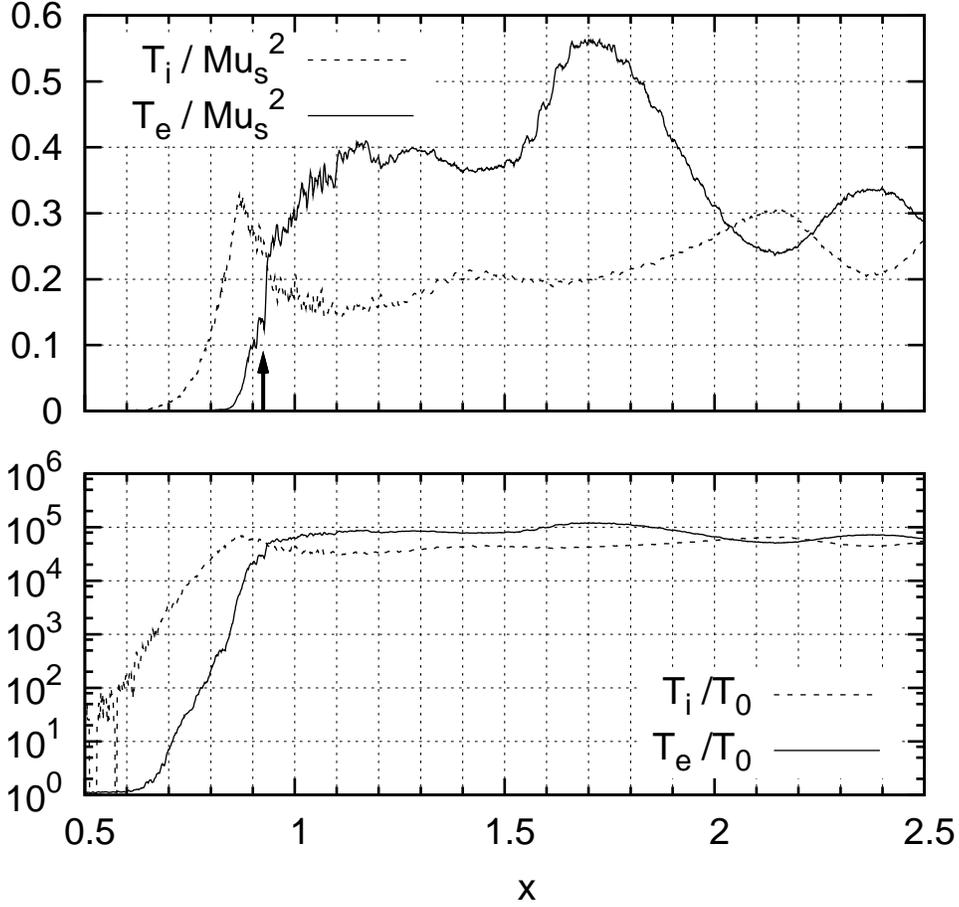}
\caption{The effective temperature variation normalized by the shocked flow energy $Mu_s ^2$ (top) and normalized by the initial temperature $T_{0}$ (bottom), where $u_s$ ($=1.5 u_0$) is an averaged injection flow speed in the shock frame. The dotted and solid lines correspond to the ion and electron temperature, respectively. The arrow in the top panel at $x \sim 0.92$ shows a position of abrupt rise of the $T_e$. This position corresponds to the boundary between region I and II discussed in the subsection III.C. The time of the figure is same with Figure 1.}
\end{figure}

\clearpage
\begin{figure} 
\includegraphics[scale=0.9]{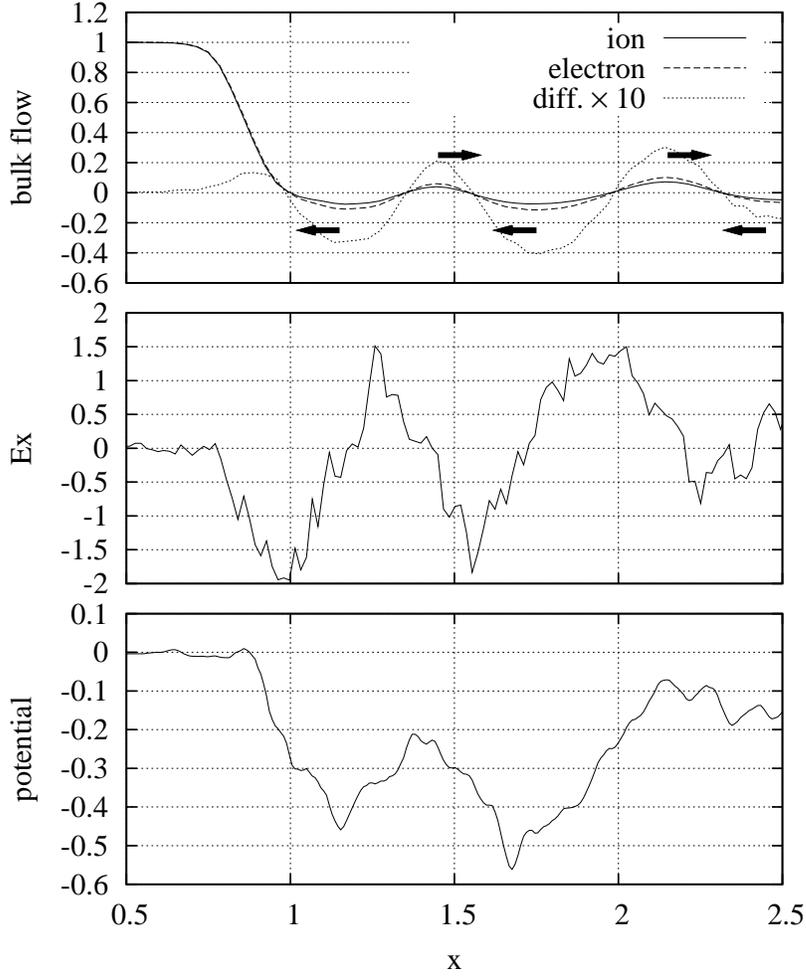}
\caption{Top: The bulk flow speed of the ion (solid), electron (dashed), and the flow speed difference of these two components (dotted, amplified 10 times) normalized by $u_0$. Arrows in the panel shows the direction of the electron component shooting out of the ion component. middle: $E_x$ profile normalized by $E_{0}$. Top and middle profiles are smoothed over 25$c/\omega_{pi}$ with each smoothing point at an interval of 2$c/\omega_{pi}$.
bottom: The electrostatic potential corresponding to the above $E_x$ profile. The data is normalized by $M u_s^2 /2e$ and smoothed over 5$c/\omega_{pi}$ with each smoothing point at an interval of $c/\omega_{pi}$. The time of the figure is same with Figure 1.}
\end{figure}

\clearpage
\begin{figure} 
\includegraphics{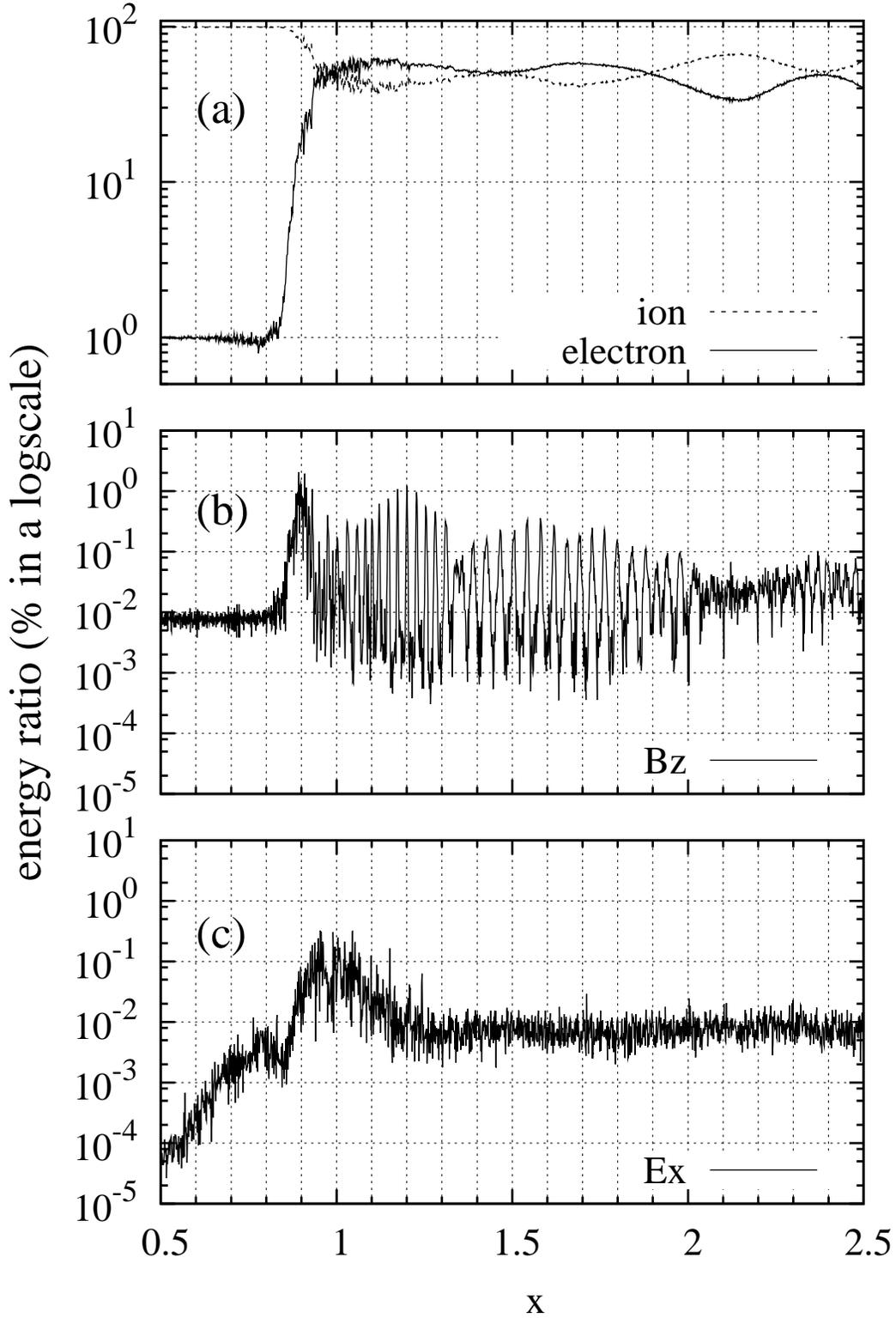}
\caption{The variation of the energy ratio (\%) among (a) the ions and electrons (dotted and solid line), (b) the magnetic field , and (c) the electric field from the upstream to the downstream side of the shock transition region ($x = 0.5 \sim 2.5$). Note the vertical axis is a log-scale. The time of the figure is same with Figure 1.}
\end{figure}

\clearpage
\begin{figure} 
\includegraphics[scale=0.8]{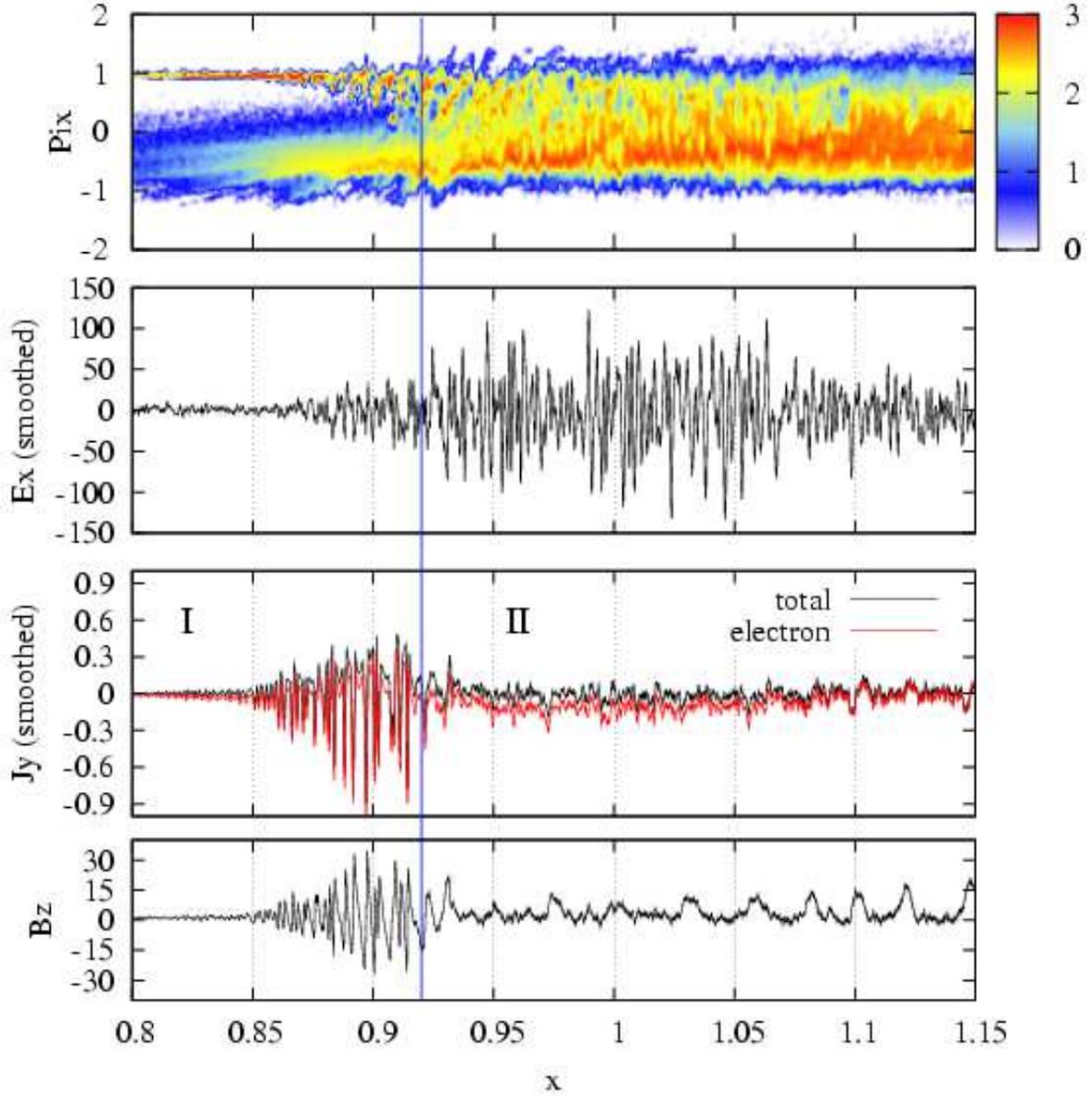}
\caption{(Color online) The enlargement picture of Figure 1, an entrance of the shock transition region. From top, ion phase space density ($P_{ix}$), $E_x$, electric current $J_y$ of the electron (the lower line) and total (the upper line), and $B_z$. Smoothed profiles ($J_y$ and $E_x$) are obtained by averaged over $c/\omega_{pe}$ with each drawing point shifted at an interval of $c/6 \omega_{pe}$. The vertical line indicates boundary between the region I and II. The time of the figure is same with Figure 1.}
\end{figure}

\clearpage
\begin{figure} 
\includegraphics[scale=0.7]{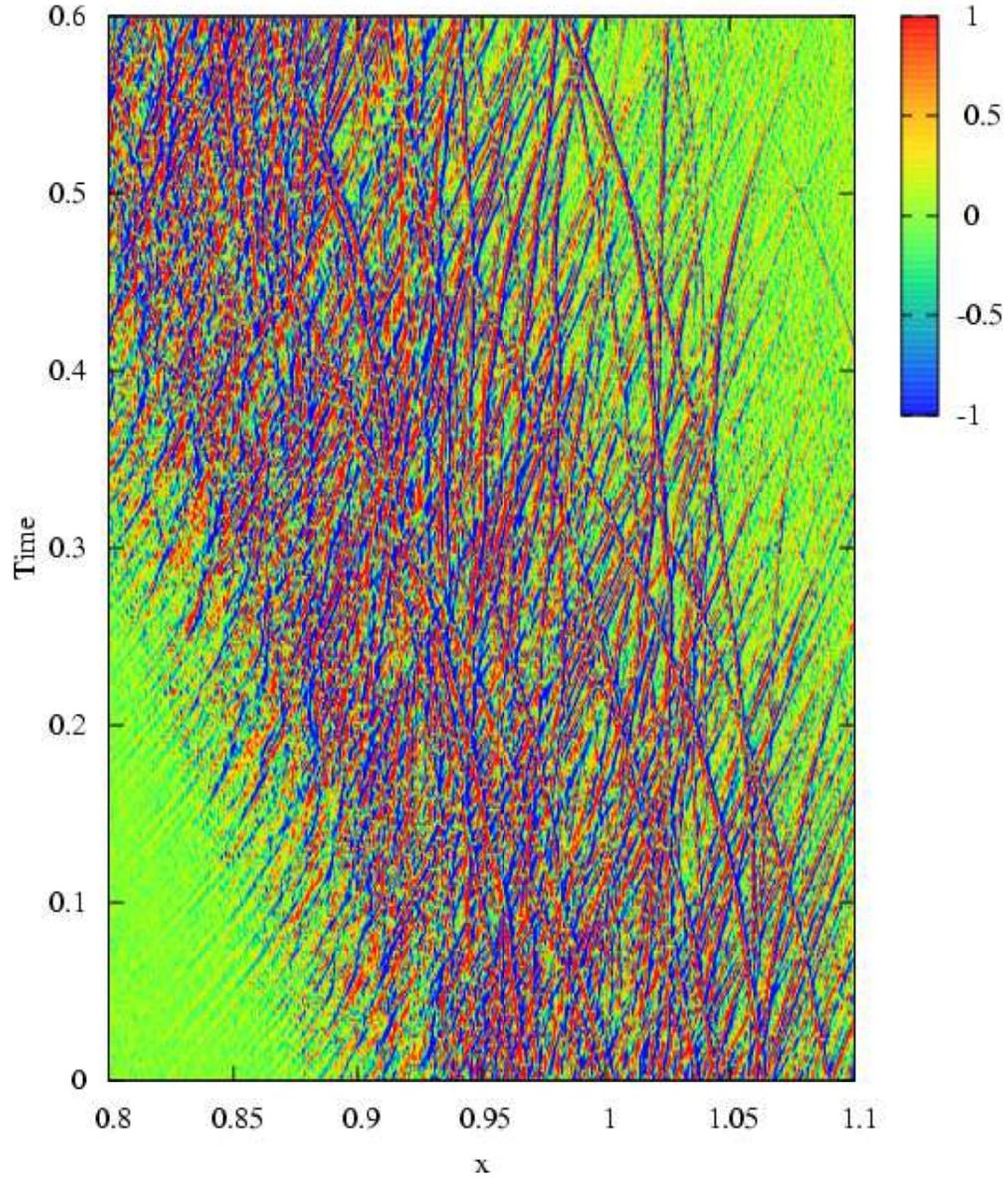}
\caption{(Color online) Trajectories of large-amplitude electric field ($E_x$) in the x-time phase space. The relative amplitude of $E_x$ is shown by the contour bar. Time is normalized by $T_{STR}$. The origin of the time in the figure is same with Figure 1.}
\end{figure}

\clearpage
\begin{figure} 
\includegraphics[scale=0.7]{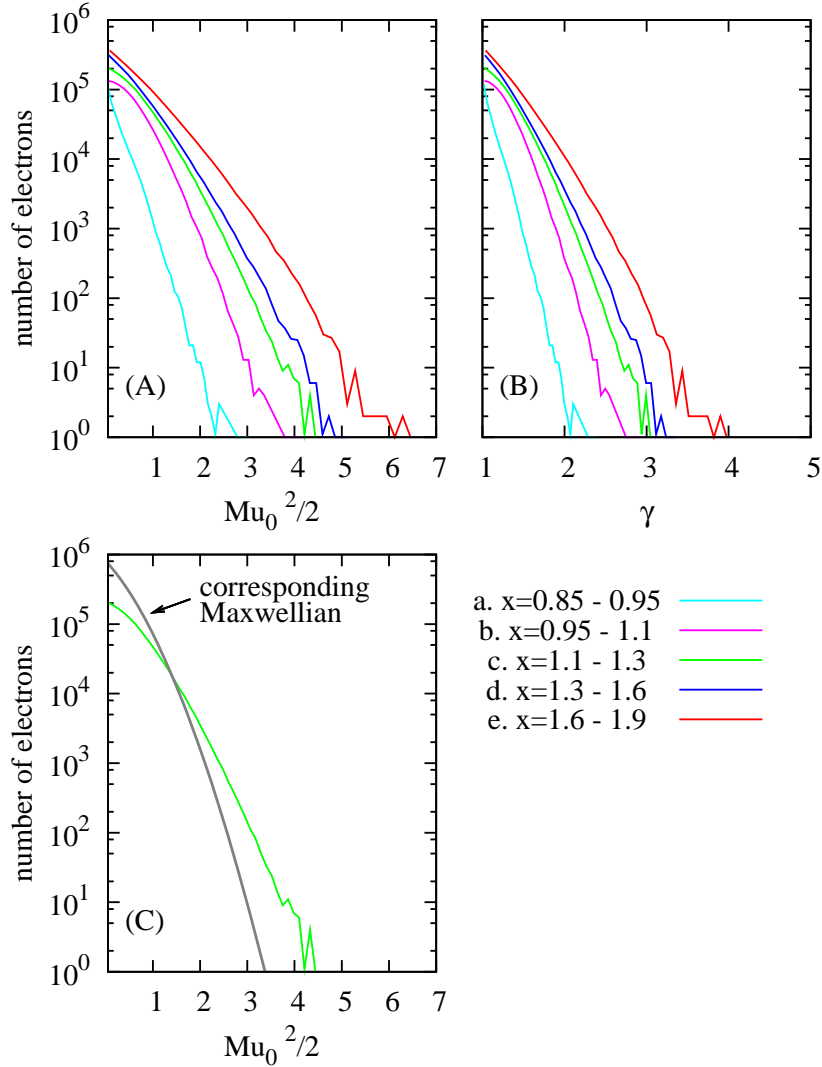}
\caption{(Color online) The electron energy spectra. The energy is normalized by the ion injection flow energy ($Mu_0 ^2/2$) in the panels (A) \& (C). Each spectrum in the panels (A) \& (B) is sampled, from left, at just entrance of the STR (a. $x = 0.85 \sim 0.95$), around the ramp region (b. $x = 0.95 \sim 1.1$), around the magnetic overshoot region (c. $x = 1.1 \sim 1.3$), around the magnetic undershoot region (d. $x = 1.3 \sim 1.6$), and its downstream region (e. $x = 1.6 \sim 1.9$). These regions corresponds to the regions indicated by arrows (a) $\sim$ (e) in Figure 1.  In the panel (C) the energy spectrum c. is drawn with a corresponding Maxwellian with the same effective temperature in this region. The time of the figure is same with Figure 1.}
\end{figure}

\clearpage
\begin{figure} 
\includegraphics[scale=0.7]{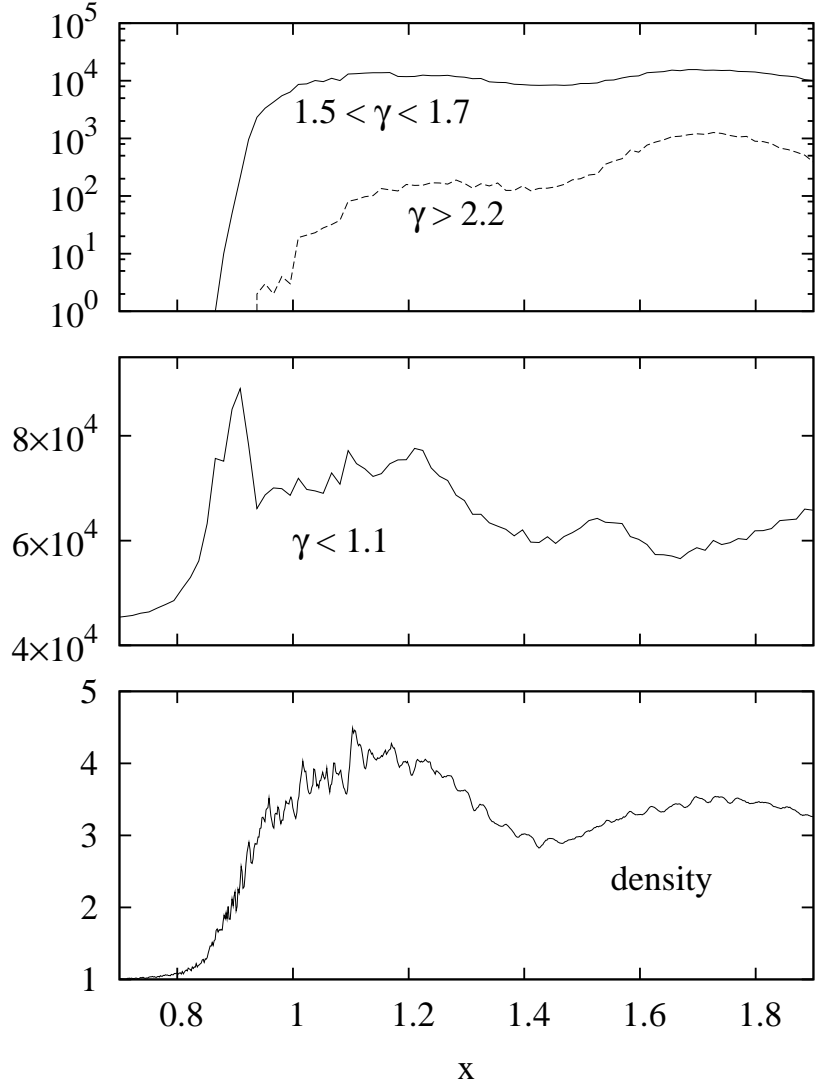}
\caption{The electron number distribution with energy $\gamma>2.2$ (top dashed line), $1.5 < \gamma < 1.7$ (top solid line), and $\gamma < 1.1$ (middle solid line). The electron density profile normalized by the upstream density is shown for reference (bottom). The density profile is smoothed by averaged over $c/\omega_{pi}$. Note the top vertical axis is a log-scale. The time of the figure is same with Figure 1.}
\end{figure}

\clearpage
\begin{figure} 
\includegraphics[scale=0.7]{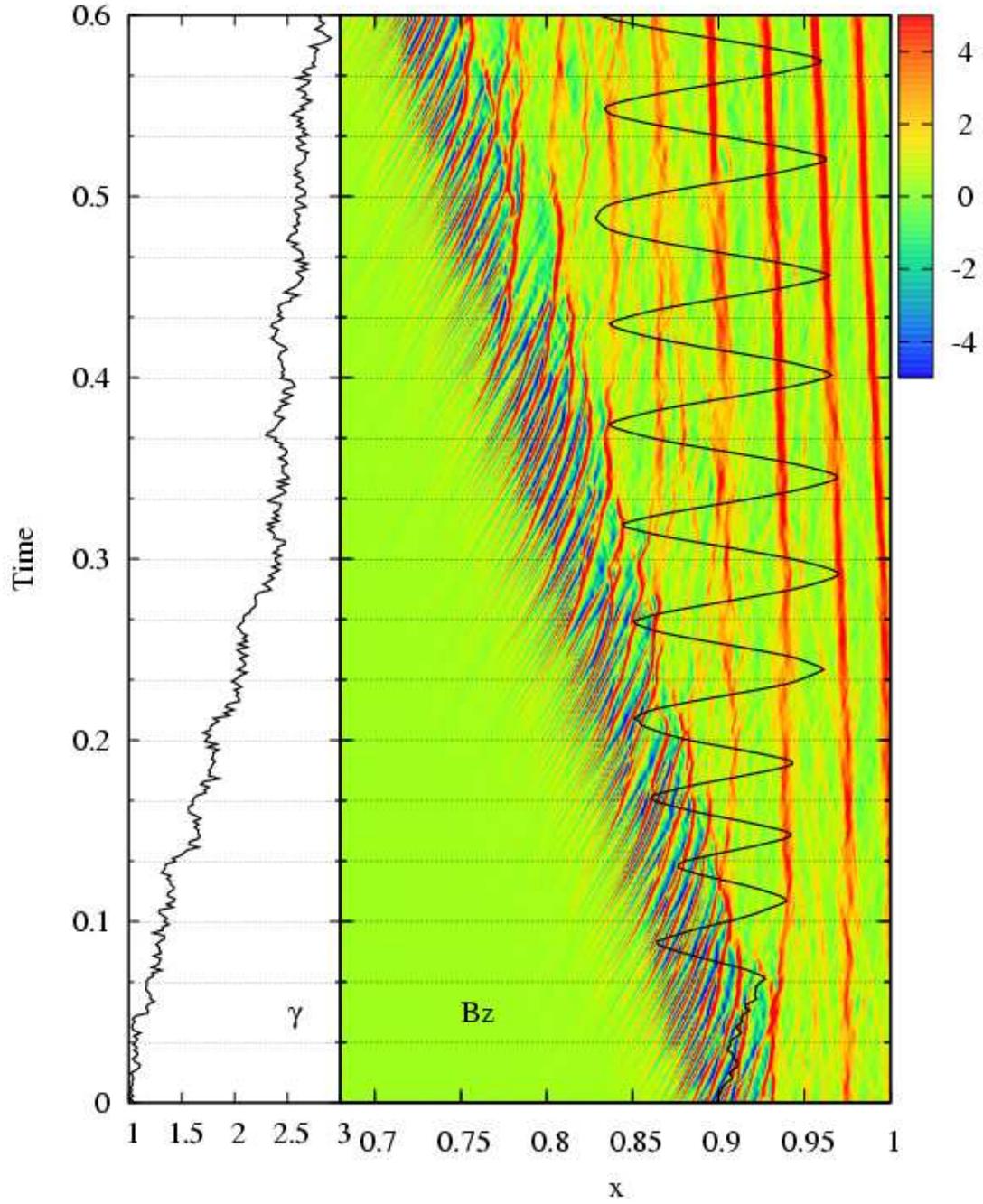}
\caption{(Color online) An example of the electron orbit around the entrance of the STR on the time-stacked profiles of the magnetic field $B_z$ (right). The strength of $B_z$ is shown by the contour The left panel shows the electron energy variation.  The origin of the time in the figure is same with Figure 1.}
\end{figure}

\clearpage
\begin{figure} 
\includegraphics[scale=0.7]{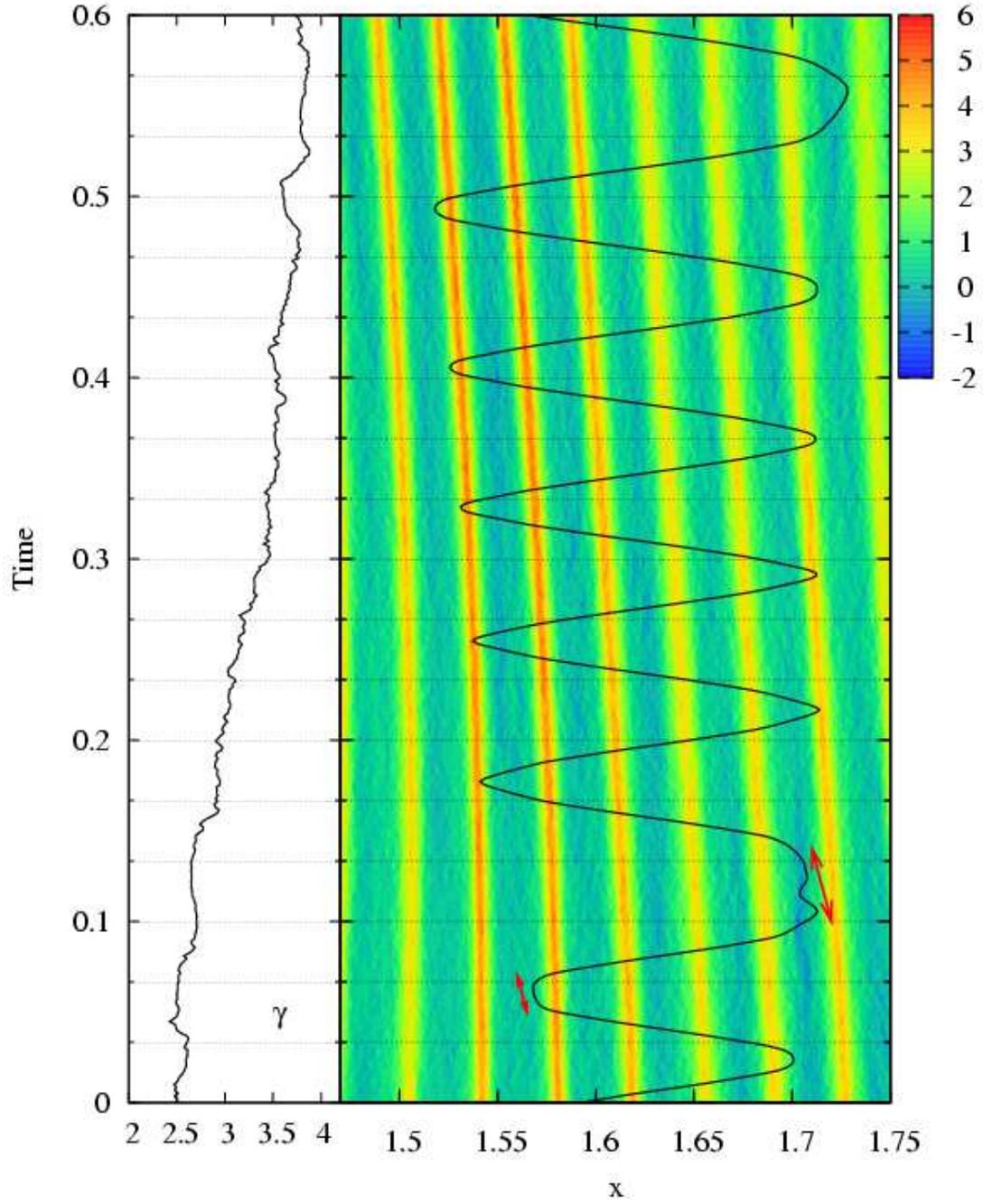}
\caption{(Color online) An example of the electron orbit and energy variation in the downstream side. The format is the same with Figure 10. Two arrows in the right panel indicate the regions where the electron shows unmagnetized behavior.  The origin of the time in the figure is same with Figure 1.}
\end{figure}

\clearpage
\begin{figure} 
\includegraphics{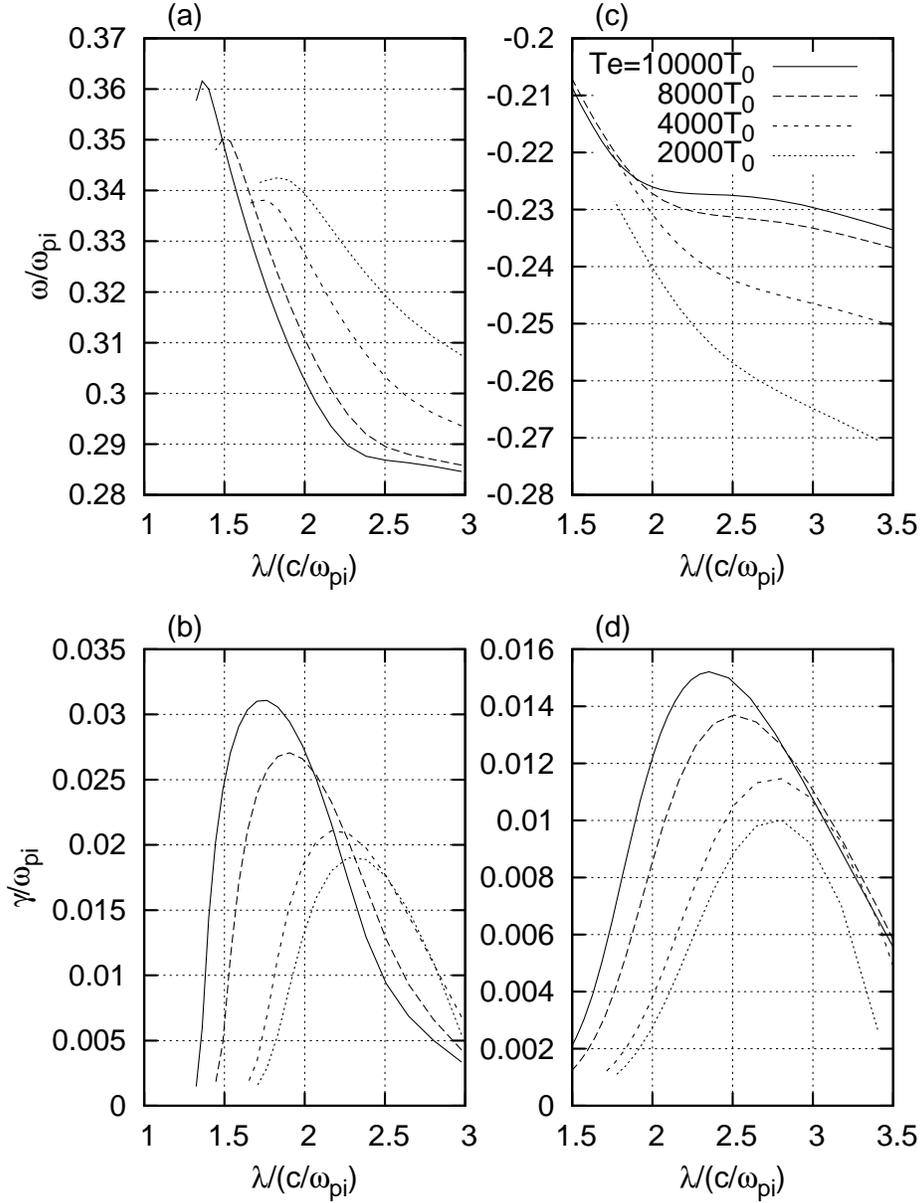}
\caption{The dispersion relation $\omega$ versus $\lambda$ (wavenumber) ((a) \& (c)) and growth rate $\gamma$ versus $\lambda$ ((b) \& (d)) of the peculiar mode in the regime of the ion-ion interaction on the incoming ion ((a) \& (b)), and on the reflected ion ((c) \& (d)). $\omega$ and $\lambda$ are normalized by the plasma frequency of the upstream ion $\omega_{pi}$ and its inertia $c/\omega_{pi}$, respectively.}
\end{figure}

\end{document}